\documentclass[11pt]{article}
\usepackage[utf8]{inputenc}
\usepackage[T1]{fontenc}
\pdfoutput = 1

\usepackage{amsmath}
\usepackage{amssymb}
\usepackage{bbm}
\usepackage{bbold}
\usepackage{braket}
\usepackage{caption}
\usepackage{color}
    \definecolor{darkgreen}{rgb}{0,0.5,0}
    \definecolor{darkred}{rgb}{0.5,0,0}
    \definecolor{darkblue}{rgb}{0,0,0.6}
    \definecolor{purple}{rgb}{0.4,.2,0.7}
\usepackage[mathcal]{eucal}
\usepackage{float}
\usepackage{graphicx}
\usepackage[hyperfootnotes = true, colorlinks = true, linkcolor = darkblue, citecolor = purple]{hyperref}
\usepackage{mathabx}
\usepackage{mathtools}
\usepackage{pdfsync}
\usepackage{slashed}
\usepackage[normalem]{ulem}
\usepackage{upgreek}
\usepackage{url}

\usepackage[margin = 2.2cm]{geometry}
\usepackage[ragged]{footmisc}
\newcommand{\p}{\partial}

\def\Re{\,{\rm Re}\,}

\def\({\left(}
\def\){\right)}
\def\[{\left[}
\def\]{\right]}

\newcommand{\qeq}{{\hbox{=\kern-2.3mm ? \kern.5mm }}}
\renewcommand{\qeq}{=}

\newcommand{\eps}{\epsilon}

\newcommand{\OO}{{\cal O}}

\newcommand{\wt}{\widetilde}

\def\cl0{\tilde c_0}

\def\one{{\hbox{ 1\kern-.8mm l}}}
\def\zero{{\hbox{ 0\kern-1.5mm 0}}}

\def\be{\begin{equation}}
\def\ee{\end{equation}}

\renewcommand{\d}{\mathrm{d}}
\renewcommand{\i}{\mathrm{i}}
\renewcommand{\tilde}{\widetilde}
\DeclareMathOperator{\Tr}{Tr}

\newcommand{\mutilde}{\widetilde{\mu}}
\newcommand{\zbar}{\overline{z}}
\newcommand{\wbar}{\overline{w}}
\newcommand{\cbar}{\overline{c}}
\mathchardef\mhyphen="2D

\numberwithin{equation}{section}

\newcommand{\ben}{\begin{eqnarray}\displaystyle}
\newcommand{\een}{\end{eqnarray}}

\newcommand{\refb}{\eqref}

\newcommand{\qt}{v}

\newcommand{\lt}{\tilde{\lambda}}
\newcommand{\zt}{\widetilde{z}}

\def\figthree{

    \def\JPicScale{0.8}
    \ifx\JPicScale\undefined\def\JPicScale{1}\fi
    \unitlength \JPicScale mm
    
    \begin{picture}(70,60)(0,0)
    
        \linethickness{1mm}
        \put(10,50){\line(1,0){20}}
        \linethickness{0.2mm}
        \put(30,50){\line(1,0){20}}
        \linethickness{1mm}
        \put(50,50){\line(1,0){20}}
        
        \linethickness{1mm}
        \put(90,50){\line(1,0){20}}
        \linethickness{1mm}
        \put(110,50){\line(1,0){20}}

        \put(30,50){\makebox(0,0)[cc]{$\times$}}
        \put(50,50){\makebox(0,0)[cc]{$\times$}}
        \put(110,50){\makebox(0,0)[cc]{$\times$}}

        \put(40,40){\makebox(0,0)[cc]{(a)}}
        \put(110,40){\makebox(0,0)[cc]{(b)}}
    \end{picture}

}

\def\figsix{
    \def\JPicScale{1}
    \ifx\JPicScale\undefined\def\JPicScale{1}\fi
    \unitlength \JPicScale mm
    
    \begin{picture}(140,55.59)(0,0)   
        \linethickness{0.2mm}
        \put(45.6,50){\circle{11.18}}
        
        \linethickness{1mm}
        \put(20,50){\line(1,0){10}}
        \linethickness{0.2mm}
        \put(30,50){\line(1,0){10}}
        
        \linethickness{1mm}
        \put(60,50){\line(1,0){10}}
        \linethickness{0.2mm}
        \put(70,50){\line(1,0){10}}
        
        \linethickness{1mm}
        \put(90,50){\line(1,0){10}}
        \linethickness{0.2mm}
        \put(105.5,50){\circle{11.18}}
        
        \linethickness{1mm}
        \put(125,50){\line(1,0){15}}
        
        \put(30,50){\makebox(0,0)[cc]{$\times$}}
        \put(40,50){\makebox(0,0)[cc]{$\times$}}
        \put(70,50){\makebox(0,0)[cc]{$\times$}}
        \put(80,50){\makebox(0,0)[cc]{$\otimes$}}
        \put(100,50){\makebox(0,0)[cc]{$\times$}}
        \put(140,50){\makebox(0,0)[cc]{$\otimes$}}
        
        \put(35,35){\makebox(0,0)[cc]{(a)}}
        \put(72,35){\makebox(0,0)[cc]{(b)}}
        \put(105,35){\makebox(0,0)[cc]{(c)}}
        \put(135,35){\makebox(0,0)[cc]{(d)}}
        
        \put(35,53){\makebox(0,0)[cc]{$q_1$}}
        \put(45,58){\makebox(0,0)[cc]{$q_2$}}
        \put(75,53){\makebox(0,0)[cc]{$q_1$}}
        \put(105,58){\makebox(0,0)[cc]{$q_2$}} 
    \end{picture}
}

\def\figcooo{

    \def\JPicScale{0.8}
    \ifx\JPicScale\undefined\def\JPicScale{1}\fi
    \unitlength \JPicScale mm
    \begin{picture}(150,70)(0,0)
    
    \linethickness{1mm}
    \put(10,50){\line(1,0){15}}
    \linethickness{0.2mm}
    \put(25,50){\line(1,0){15}}
    \linethickness{0.2mm}
    \multiput(40,50)(0.12,-0.12){83}{\line(1,0){0.12}}
    \linethickness{0.2mm}
    \multiput(40,50)(0.12,0.12){83}{\line(1,0){0.12}}
    \linethickness{0.2mm}
    \multiput(50,60)(0.12,0.24){42}{\line(0,1){0.24}}
    \linethickness{0.2mm}
    \put(50,60){\line(1,0){10}}
    
    \linethickness{1mm}
    \put(70,50){\line(1,0){10}}
    \linethickness{0.2mm}
    \put(80,50){\line(1,0){10}}
    \linethickness{0.2mm}
    \multiput(90,50)(0.12,0.12){83}{\line(1,0){0.12}}
    \linethickness{0.2mm}
    \put(90,50){\line(1,0){10}}
    \linethickness{0.2mm}
    \multiput(90,50)(0.12,-0.12){83}{\line(1,0){0.12}}
    
    \linethickness{1mm}
    \put(115,50){\line(1,0){15}}
    \linethickness{0.2mm}
    \multiput(130,50)(0.12,0.12){83}{\line(1,0){0.12}}
    \linethickness{0.2mm}
    \multiput(130,50)(0.12,-0.12){83}{\line(1,0){0.12}}
    \linethickness{0.2mm}
    \multiput(140,60)(0.12,0.24){42}{\line(0,1){0.24}}
    \linethickness{0.2mm}
    \put(140,60){\line(1,0){10}}
    
    \linethickness{1mm}
    \put(165,50){\line(1,0){15}}
    \linethickness{0.2mm}
    \multiput(180,50)(0.12,0.12){83}{\line(1,0){0.12}}
    \linethickness{0.2mm}
    \put(180,50){\line(1,0){15}}
    \linethickness{0.2mm}
    \multiput(180,50)(0.12,-0.12){83}{\line(1,0){0.12}}
    
    \put(25,30){\makebox(0,0)[cc]{(a)}}
    \put(85,30){\makebox(0,0)[cc]{(b)}}
    \put(135,30){\makebox(0,0)[cc]{(c)}}
    \put(180,30){\makebox(0,0)[cc]{(d)}}

    \put(32,53){\makebox(0,0)[cc]{$q_1$}}
    \put(43,58){\makebox(0,0)[cc]{$q_2$}}
    \put(86,53){\makebox(0,0)[cc]{$q_1$}}
    \put(134,59){\makebox(0,0)[cc]{$q_2$}}
    
    \end{picture}
}

\def\figpsione{
    
    \def\JPicScale{1.2}
    \ifx\JPicScale\undefined\def\JPicScale{1}\fi
    \unitlength \JPicScale mm
    \begin{picture}(140,55.59)(0,0)
        \linethickness{0.2mm}
        \put(45.6,50){\circle{11.18}}
        
        \linethickness{1mm}
        \put(20,50){\line(1,0){10}}
        \linethickness{0.2mm}
        \put(30,50){\line(1,0){10}}

        \linethickness{1mm}
        \put(90,50){\line(1,0){10}}
        \linethickness{0.2mm}
        \put(105.5,50){\circle{11.18}}

        \put(30,50){\makebox(0,0)[cc]{$\times$}}        
        \put(100,50){\makebox(0,0)[cc]{$\times$}}
        
        \put(35,35){\makebox(0,0)[cc]{(a)}}
        \put(105,35){\makebox(0,0)[cc]{(b)}}

        \put(40,50){\makebox(0,0)[cc]{$\times$}}
        \put(55,50){\makebox(0,0)[cc]{$\psi$}}
        \put(115,50){\makebox(0,0)[cc]{$\psi$}}

    \end{picture}

}

\begin{document}

\thispagestyle{empty}
\begin{center}
    ~\vspace{5mm}
    
    {\Large \bf
    
    The ZZ annulus one-point function in non-critical string theory: \\
    A string field theory analysis
    
    }
    
    \vspace{0.4in}
    
    {\bf 
    Dan Stefan Eniceicu,$^1$ 
    Raghu Mahajan,$^1$
    Pronobesh Maity,$^{2}$
    Chitraang Murdia,$^{3,4}$ and 
    Ashoke Sen$^{2}$
    }

    \vspace{0.4in}

    $^1$ Department of Physics, Stanford University, Stanford, CA 94305, USA \vskip1ex
    $^2$ International Centre for Theoretical Sciences, 
    Bengaluru - 560089, India \vskip1ex
    $^3$ Berkeley Center for Theoretical Physics, Department of Physics, University of California, Berkeley, CA 94720, USA \vskip1ex
    $^4$ Theoretical Physics Group, Lawrence Berkeley National Laboratory, Berkeley, CA 94720, USA 
    \vspace{0.1in}
    
    {\tt eniceicu@stanford.edu, raghumahajan@stanford.edu, pronobesh.maity@icts.res.in, murdia@berkeley.edu,
    ashoke.sen@icts.res.in}
\end{center}

\vspace{0.4in}

\begin{abstract}

We compute the ZZ annulus one-point function of the cosmological constant operator in non-critical string theory, regulating divergences from the boundaries of moduli space using string field theory.
We identify a subtle issue in a previous analysis of these divergences, which was done in the context of the $c=1$ string theory,  and where it had led to a mismatch with the prediction from the dual matrix quantum mechanics.
After fixing this issue, we find a precise match to the expected answer in both the $c<1$ and $c=1$ cases.
We also compute the disk two-point function, which is a quantity of the same order, 
and show that it too matches with the general prediction.

\end{abstract}

\pagebreak

\tableofcontents

\section{Introduction and summary} \label{sintro}
The study of non-perturbative effects due to ZZ instantons \cite{Zamolodchikov:2001ah} in two-dimensional string theory by Balthazar, Rodriguez, and Yin \cite{Balthazar:2019rnh} has motivated a string field theory analysis of IR divergences in instanton amplitudes.
The agreement between the string field theory analyses and the predictions from the dual matrix quantum mechanics is impressive \cite{Sen:2019qqg, Sen:2020eck, Sen:2021qdk}.
The string field theory analysis of instanton amplitudes has been extended to other non-critical string models \cite{Eniceicu:2022nay, Eniceicu:2022dru, Balthazar:2022apu, Chakravarty:2022cgj, Sen:2022clw} where the computations agree precisely with the predictions from the dual matrix models. 
It has also been extended to critical superstrings \cite{Agmon:2022vdj, Sen:2021tpp, Sen:2021jbr, Alexandrov:2021shf, Alexandrov:2021dyl} where the D-instanton effects match precisely with the predictions from superstring dualities.
In \cite{Alexandrov:2022mmy},  worldsheet computations of instanton effects were performed in Calabi-Yau orientifold compactifications, which is a new result and was not previously known from a dual description.

Among all these successes, there is one particular observable that stands out and does not match with the dual prediction.
This is the annulus one-point amplitude in the original $c=1$ string theory computation of \cite{Balthazar:2019rnh}, which is relevant for
computing the first subleading-in-$g_s$ correction to the D-instanton induced $n$-point amplitude of closed-string operators.
This amplitude receives divergent contributions from integration over the worldsheet moduli near the boundaries of moduli space.
This leads to an additive ambiguous term in the amplitude.
Extracting the finite part of the worldsheet amplitude via numerical integration over the moduli space, and comparing this with the amplitude from the dual matrix quantum mechanics leads to a prediction for the ambiguous piece of the worldsheet amplitude \cite{Balthazar:2019rnh, BRYunpublished}. 
In \cite{Sen:2020eck}, a string field theory analysis was performed to determine the ambiguous piece, but the result was found to not match with \cite{Balthazar:2019rnh, BRYunpublished}, leading to a puzzle.

In the present work, we resolve this mismatch. 
We first simplify the model by working with the $c<1$ non-critical string, and we study the integrated correlation functions of the cosmological constant operator.
The first simplification is that one does not have to worry about the translation zero mode of the $c=1$ scalar.
Second, the one-point annulus amplitude of the cosmological operator has a simple form that can be obtained by differentiating the partition function with respect to 
the world-sheet cosmological constant $\mu$.
Third, because of the Liouville equation of motion, the cosmological operator is a total derivative and the moduli space integral reduces to just boundary contributions, obviating the need for numerical integration over the moduli space.
In trying to compute the one-point annulus amplitude in this simpler model, we were able to identify a subtle issue in the computation in appendix D of \cite{Sen:2020eck}, which analyzed the disk amplitude with one closed string puncture and three open string punctures.
This was needed for finding the relation between the string field theory gauge transformation parameter and the rigid $U(1)$ transformation parameter under which an open string with one end on the instanton picks up a phase.

Let us briefly explain the subtlety. 
The disk amplitude with one closed string puncture and three open string punctures has a two-dimensional moduli space.
String field theory instructs us to integrate a two-form on a subset $S$ of this moduli space that excludes certain regions around the boundaries. 
The two-form turns out to be exact.
Let's denote it by $\d {J}$, so that the moduli-space integral reduces to the integral of ${J}$ over the boundary $\partial S$ of $S$.
It so happens that in appendix D of \cite{Sen:2020eck}, when we go once around $\partial S$, the open string punctures do not return to their original position, but to a configuration related to the initial one by the one-parameter subgroup of PSL$(2,\mathbb{R})$ that keeps the point $z=\i$ in the upper half plane fixed.
In such a situation, one must make sure that the contraction of ${J}$ with the tangent vector along the orbits of this PSL$(2,\mathbb{R})$ transformation is zero, something that was not true for the ${J}$ chosen in \cite{Sen:2020eck}.
It turns that one can add an exact one-form to ${J}$ so that it satisfies the desired property. 
(Alternatively, one can reduce the orbits under discussion to points by
fixing the location of one of the punctures and work directly with the two-dimensional moduli space.)
After this fix, the mismatch goes away and one-finds agreement with predictions from the dual matrix models both in the $c<1$ and the $c=1$ case.

Now we present the main result of this work.
Let $V = e^{2b\phi}$ denote the bulk cosmological constant vertex operator in minimal string theory. 
The integrated correlation functions of $V$ can be obtained by taking $\mu$-derivatives of the partition function.
Let $g_s^{-1}$ be the tension of the ZZ brane.
Further, let $g_s f$ denote the disk two-point of $V$, divided by the square of the disk one-point function of $V$,
and let $g_s g$ denote the annulus one-point of $V$, divided by the disk one-point function of $V$.
By taking $\mu$-derivatives of the partition function and setting $\mu = \frac{1}{\pi}$, one finds
\begin{align}
    f = \frac{2b}{Q} - 1 \, , \quad g = \frac{1}{2}\, .
    \label{fg_intro}
\end{align}
Our goal will be to reproduce both of these results via explicit integrals over the relevant moduli spaces, using string field theory to regulate divergences from the boundaries.

The rest of the paper is devoted to setting up the problem in the $c<1$ case and explaining the above remarks in more detail.
In section \ref{sec:conventions}, we briefly review the $c<1$ non-critical string theory, and also certain conventions for the Liouville and ghost CFTs and for string amplitudes that will be important for us.
In section \ref{sec:generalanalysis}, we provide the general analysis that leads to a concrete prediction for the disk two-point function and the annulus one-point function of the cosmological constant operator, leading to the predictions (\ref{fg_intro}).
The key point is that the correlation functions of the cosmological constant operator can be related to $\mu$-derivatives of the partition function.
In section \ref{sec:disk2pt}, we compute the disk two-point function of the cosmological constant operator by integrating over the one-dimensional moduli space, which matches with the prediction (\ref{fg_intro}).
The disk two-point function contributes at the same order as the annulus one-point function, and was already found to match the matrix quantum mechanics result in the $c=1$ case \cite{Balthazar:2019rnh, BRYunpublished, Sen:2020eck}.
Finally, in section \ref{sec:annulus1pt}, we compute the annulus one-point function exploiting the total derivative nature of the cosmological constant operator to integrate over the two-dimensional moduli space.
The subtlety in appendix D of \cite{Sen:2020eck} is explained in section \ref{sec:gghost}.
The analysis of section \ref{sec:disk2pt} and section \ref{sec:annulus1pt} requires computing string field theory Feynman diagrams to get finite results.
Since the analysis of \cite{Sen:2020eck} was quite lengthy overall, we will not repeat all the computational details of the various contributions and instead emphasize the conceptual points that are different in our analysis.
The appendices contain details of various overall normalizations of string amplitudes that are important for our work.

\section{Setup and conventions}
\label{sec:conventions}

\subsection{Conventions for string amplitudes}
\label{sec:ghostconventions}

We will follow the conventions of \cite{Sen:2021tpp}.
One important convention is that the integrated closed string vertex operators are integrated with the measure $\frac{\d x \, \d y}{\pi}$.
For the upper half plane geometry, the open string punctures are integrated along the real axis with measure $\d x$.

We take the three-point function of the $c$-ghost in the upper half plane to be
\begin{align}
    \langle c(z_1) c(z_2) c(z_3)\rangle_\text{UHP} = - (z_1-z_2) (z_2-z_3) (z_1-z_3) \, .
    \label{cghostcorrelator}
\end{align}
This normalization, with the string field theory path integral being weighted as $\exp [ \frac{1}{2}\, \langle \Psi|Q_B|\Psi\rangle ]$, gives rise to the path integral weight $\exp[ -\frac{1}{2} h_b \phi_b^2]$ for a Siegel-gauge bosonic field $\phi_b$ with $L_0 = h_b$.\footnote{In our convention where the path integral is weighted by exponential of the action,  an $n$-point interaction term in the action gives a contribution to the $n$-point amplitude without any extra minus sign or factor of $\i$.}
For example, in the case of the the tachyon $\phi_1 c_1|0\rangle$, the action evaluates to
$\frac{1}{2} \phi_1^2 \, \langle 0|c_{-1} \, c_0L_0 \, c_1|0\rangle = \frac{1}{2} \phi_1^2$, which is the correct result since $h_b=-1$ for the tachyon.
For later use, we also need the out-of-Siegel gauge field $\psi$ that appears in the string field as
\begin{align}
    \Psi = \phi_1 c_1 \ket{0} + \i \psi c_0\ket{0} + \ldots 
\end{align}
The path integral weight of $\psi$ is $\exp(-\psi^2)$, so that the propagator of $\psi$ equals $\frac{1}{2}$.

We define $g_s$ so that nonperturbative contributions to string amplitudes carry an overall factor of $\exp(-g_s^{-1})$.  
In other words, the action of the instanton is $-g_s^{-1}$.


In the conventions of \cite{Sen:2021tpp},  the one point function of a closed string vertex operator $\psi_c$ on the upper half plane is given by
\begin{align}
    A_\text{disk}(\psi_c) = 
    \frac{1}{4g_s} \left \langle (\partial c - \overline{\partial}\cbar)
    \psi_c \right \rangle_\text{UHP}\, .
    \label{adiskv}
\end{align}
The upper half plane amplitude for $n$ closed string punctures and $m$ open string punctures, with one closed puncture and one open string puncture fixed is
\begin{align}
    A_\text{disk}(\psi_c^n \psi_o^m) = \frac{\i \pi}{g_s}
    \int \langle \psi_c^n \psi_o^m \rangle_\text{UHP}\, .
    \label{fix1c1o}
\end{align}
The factor of $\i$ was explained in appendix A of \cite{Alexandrov:2021shf}.
In this paper, we will instead be interested in the case where the PSL$(2,\mathbb{R})$ symmetry is fixed by fixing the position of one closed string puncture at $z= \i$ and fixing the $x$-coordinate of another closed string puncture to zero.
In appendix \ref{app:fixtwoclosedstrings} we show that the amplitude in this gauge fixing takes the form
\begin{align}
    A_\text{disk}(\psi_c^n \psi_o^m) = \frac{\i}{2 g_s} 
    \int \langle \psi_c^n \psi_o^m \rangle_\text{UHP}\, ,
    \label{fixtwoclosed}
\end{align}
where one closed string puncture is fixed at $z = \i$ and takes the form $c \overline{c} V(\i)$, and a second closed string puncture is integrated with measure $\d y$ from $y=0$ to $y=1$ and takes the form $(c + \overline{c}) V(\i y)$.
The important point about (\ref{fixtwoclosed}) is the precise overall numerical factor.

\subsection{Minimal string theory and Liouville CFT}
The term minimal string theory refers to a worldsheet model where the matter sector consists of the $(p',p)$ minimal model \cite{DiFrancesco:1993cyw, Seiberg:2004at}. 
Here $p'$ and $p$ are two relatively prime integers and our convention is that $p' < p$.
The minimal model is a CFT with central charge $c = 1 - \frac{6(p-p')^2}{pp'}$.
The conformal mode of the metric does not decouple and gives rise to the Liouville CFT with central charge $26 - c$.
Together with the $bc$-ghosts we have an anomaly free worldsheet theory.
These theories have a dual description via an integral over two Hermitian matrices in the double-scaling limit \cite{Douglas:1990pt, Daul:1993bg}.
When $p'=2$, the integral over one of the matrices is purely Gaussian and it can be integrated out giving rise to a one-matrix integral.

The path integral of Liouville field theory on a two-dimensional Euclidean manifold with metric $g$ is given by \cite{Zamolodchikov:2001ah}
\begin{align}
    \int [D\phi] \exp\left[ - 
    \int \d x\, \d y \, \sqrt{g}\right.  & \left. \left( \frac{1}{4\pi} \, g^{\mu\nu} \partial_\mu\phi \partial_\nu \phi + \frac{1}{4\pi} Q R \phi + \mu \, e^{2 b \phi} \right) \right] \, , \quad \text{where} 
    \label{eq:Liouvilleaction}\\
 Q = \frac{1}{b} + b \, , &\quad b = \sqrt{\frac{p'}{p}}\, .
\end{align}

We will be interested in the correlation function of the marginal operator $V := e^{2 b \phi}$.
Fixing the background metric to be flat, one sees that the equation of motion of the Liouville field is
\begin{align}
    \frac{1}{4\pi} (\partial_x^2 + \partial_y^2 ) \phi = \mu\, b \, e^{2 b \phi}\, .
    \label{liouville_eom}
\end{align}
So we see that the operator $e^{2 b \phi}$ is a total derivative.
While this derivation holds in the semi-classical limit $b \to 0$, the result holds even for finite $b$ \cite{Zamolodchikov:2003yb}.
In fact, this equation of motion is only the first in an infinite series of ``higher'' equations of motion in Liouville theory \cite{Zamolodchikov:2003yb}.
These equations of motion have been previously used to compute integrated correlation functions in minimal string theory on the sphere topology \cite{zthree, bzcollection, Belavin:2008kv, Artemev:2022rng}.
It will be useful for us to introduce a rescaled cosmological constant
\begin{align}
    \widetilde{\mu} := \pi \mu 
\end{align}
so that the interaction term in the Liouville action takes the form $\int \frac{\d x \d y}{\pi}\, \widetilde{\mu}\, e^{2 b \phi}$.
This will be useful, since we will be integrating closed string punctures with the measure $\frac{\d x \d y}{\pi}$, as in \cite{Sen:2021tpp}.
With the definition of $\widetilde{\mu}$ and changing to complex coordinates $z=x+\i y$, we rewrite (\ref{liouville_eom}) as
\begin{align}
    V = e^{2 b \phi} &= \frac{1}{\widetilde{\mu} \, b} \, \partial \overline{\partial} \phi \, .
    \label{veom}
\end{align}

Another fact that we will need is the following. 
Because of the term in the action proportional to $Q R \phi$, the operators $\partial \phi$ and $\overline{\partial} \phi$ are not conformal primaries but transform as
\begin{align}
    \label{eftrs1}
    \p\phi(z, \zbar) \to {Q\over 2} {f''(z)\over f'(z)} + f'(z) \, \p\phi(f(z), \overline{f(z)})\, ,
\end{align}
with a similar relation for $\overline{\partial}\phi$.
We will also need the following OPE between the operators $\partial \phi$ and $V$
\begin{align}
    \partial \phi(z, \zbar) V(\i) = - \frac{b}{z - \i} \, V(\i) + \ldots \, .
    \label{delphiVope}
\end{align}

Instanton effects in minimal string theory are represented by open string worldsheets with ZZ boundary conditions, which are labeled by a pair of integers \cite{Zamolodchikov:2001ah}.
In this paper we will only work with one ZZ brane of the simplest $(1,1)$ type.
The boundary state for the matter sector will be taken to be the Cardy state that contains only the identity operator in the open string channel \cite{Cardy89, Cardy2004}.
Denoting by $2\pi t$ the Euclidean time in the open string channel and letting 
\begin{align}
    v := e^{-2\pi t} \, ,
    \label{vdef}
\end{align}
the annulus partition function of the minimal string with the desired boundary conditions is given by \cite{Zamolodchikov:2001ah, Rocha-Caridi, Martinec:2003ka}
\begin{align}
    C_2 &:= \int_0^\infty \frac{\d t}{2t} \, Z(v) \, , \quad \text{with} \label{eq:emptyannulus}\\
    Z(v) &= (v^{-1} - 1) \, v^{-(p-p')^2/4pp'}  \sum_{j=-\infty}^{\infty}
    [ v^{(2pp' j + p - p')^2/4pp'} - v^{(2pp' j + p + p')^2/4pp'} ] \label{zofv} \\
    &= v^{-1} - 2 + O(v) \, . \label{annulus-small-v}
\end{align}
In the last line we have displayed the terms that give rise to divergent contributions from the $t = \infty$ end of the integral.
The $v^{-1}$ term arises from the open string tachyon that multiplies the state $c_1 \vert 0 \rangle$ with $L_0 = -1$, and the $-2$ arises from two Grassmann-odd modes $p_1$ and $q_1$ that multiply the states $\vert 0 \rangle$ and $c_{-1} c_{1} \ket{0}$ with $L_0 = 0$.
One of the insights of \cite{Sen:2019qqg} was that $p_1$ and $q_1$ are ghost zero modes that arise because of the breakdown of Siegel gauge. 
This is related to the fact that the worldvolume of a D-instanton is zero dimensional.
One needs to instead integrate over the ghost-number one field $\psi$ that multiplies the state $c_0\ket{0}$ and divide by the volume of the ``gauge group'' of the worldvolume theory of the D-instanton, which is finite:
\begin{align}
    \int \d p_1 \d q_1 \quad \to \quad \frac{\int \d \psi \exp(-\psi^2)}{\int \d\theta} \, .
    \label{pq_to_psi}
\end{align}
In particular, the open string field $\psi$ can run in internal propagators; such contributions to string amplitudes are not captured by the worldsheet analysis and need to be explicitly added \cite{Sen:2020eck}.

Let us now discuss the bulk-boundary OPE with ZZ boundary conditions.
Working with the upper half plane coordinate system, this takes the form \cite{Zamolodchikov:2001ah}
\begin{align}
    \partial \phi (z, \zbar) &= - \frac{Q}{z-\zbar} + O(z-\zbar) \label{bulkbdryope1} \\
    \overline{\partial} \phi (z, \zbar) &= + \frac{Q}{z-\zbar} + O(z-\zbar) \, . \label{bulkbdryope2}
\end{align}
The coefficient of the leading term is fixed by using the equation of motion (\ref{veom}) and the fact that the one-point function of $V(z,\zbar)$ is given by $\frac{Q}{\mutilde b} \frac{1}{\vert z - \zbar \vert^2}$ \cite{Zamolodchikov:2001ah}.
Setting $\mutilde = 1$, note in particular that 
\begin{align}
\langle V(\i) \rangle_\text{UHP} = \frac{Q}{4b}.  
\label{vonepointi}
\end{align}
That the expression for the one-point function of $V(z,\zbar)$ resembles the metric of hyperbolic space $\mathbb{H}^2$ (represented using the upper half plane) and with the correct radius of curvature in the semi-classical limit is tied to the fact that physically the ZZ boundary conditions represent the Liouville field theory placed on the pseudosphere.
Note also that there are no $O((z - \zbar)^0)$ terms in (\ref{bulkbdryope1}) and (\ref{bulkbdryope2}) since the boundary theory in the Liouville sector does not have any state of conformal weight 1 (the only boundary operator is the identity, so that the $L_{-1}$ descendant is null).
Finally,  we note that (\ref{eftrs1}), (\ref{bulkbdryope1}), (\ref{bulkbdryope2}) and the scale invariance of the upper half plane imply that
\begin{align}
\langle \partial\phi(z,\zbar) \rangle_\text{UHP} = - \frac{Q}{z-\zbar} \, , 
\quad 
\langle \overline{\partial}\phi(z,\zbar) \rangle_\text{UHP} =  \frac{Q}{z-\zbar} \, .
\label{ephiuhp}
\end{align}

\section{General predictions for the disk two-point function and the annulus one-point function}
\label{sec:generalanalysis}
In perturbation theory, the string partition function is given by a sum over closed string worldsheets, organized by genus.
This series is asymptotic and contains non-perturbative corrections due to instanton effects \cite{Shenker:1990uf, Ginsparg:1991ws, David:1992za}, which, in the case of critical strings can be thought as being due to worldsheets with Dirichlet boundaries in all target space directions \cite{Polchinski:1994fq}.
In the context of minimal strings, the relevant boundary conditions are of the ZZ type \cite{Zamolodchikov:2001ah}, which are analogous to Dirichlet boundary conditions in the critical string.
As already mentioned, throughout this paper we will focus on a single ZZ brane of the simplest $(1,1)$ type, and also the simplest Cardy state for the matter sector.

The string partition function including the one-instanton contribution is given by
\begin{align}
\mathcal{Z}(\vec t, g_s) &= \mathcal{Z}^{(0)}(\vec t, g_s) + \mathcal{Z}^{(1)}(\vec t, g_s) + \ldots \, , \label{z0z1sum}\\
\mathcal{Z}^{(1)}(\vec t, g_s) &= \mathcal{Z}^{(0)}(\vec t, g_s)\exp\left[g_s^{-1} A(\vec t) + {1\over 2}\log g_s + B(\vec t) + C(\vec t) g_s +\cdots\right]\, ,
\label{eaa0}
\end{align}
where $g_s$ is the closed string coupling, $\vec t$ are the set of parameters that label the various possible deformations of minimal string theory or the dual double-scaled matrix model,  and $\mathcal{Z}^{(0)}(\vec t, g_s)$ is the perturbative contribution to the partition function. The dots in (\ref{z0z1sum}) denote contributions from multi-instantons,  which we do not study in this paper.\footnote{Even though the $g_s$ dependence of $\mathcal{Z}(\vec t, g_s)$ can be determined from its $\vec{t}$ dependence,  we regard $g_s$ as an independent variable.}
The point $\vec{t} = \vec{0}$ represents the conformal background, with only the worldsheet cosmological constant switched on \cite{Moore:1991ir}. 
The coefficient $\frac{1}{2}$ multiplying the $\log g_s$ term is special to the $c<1$ minimal string \cite{Eniceicu:2022nay}.

Taking derivatives of $\mathcal{Z}^{(1)}(\vec t, g_s)$ with respect to the $t_i$'s, we can get the one-instanton contribution to the $n$-point function. 
The full diagrammatics of the one-instanton contribution was discussed in section 2 of \cite{Sen:2020cef}.
In taking the derivatives,  there will be terms in which one or more derivatives hit the $\mathcal{Z}^{(0)}(\vec t, g_s)$ factor in (\ref{eaa0}).
Such terms will produce closed-string worldsheet components without boundaries and will not be the subject of interest in our work,  so we shall not write them.
So we have
\begin{align}
\frac{1}{\mathcal{Z}^{(0)}} 
\frac{\partial^n \mathcal{Z}^{(1)}}{\partial t_{i_1}\cdots \partial t_{i_n}}
\bigg|_{\vec t=0} 
&\supset
e^{g_s^{-1} A(\vec 0) + {1\over 2}\log g_s + B(\vec 0) } \, g_s^{-n}
\left\{\prod_{\alpha=1}^n {\p A\over \p t_{i_\alpha}} \right\} \nonumber \\ 
& \hspace{-0.4in} \times\, \left[  1 +
g_s \sum_{\beta, \gamma=1\atop \beta<\gamma}^n { \p^2 A\over \p t_{i_\beta} \p_{t_{i_\gamma}}}
\bigg / \left( {\p A\over \p t_{i_\beta}} {\p A\over \p t_{i_\gamma}}
\right) + g_s \sum_{\beta=1}^n {\p B\over \p t_{i_\beta}} \bigg/ {\p A\over \p t_{i_\beta}}
+ g_s \, C
+ O(g_s^2) \right ] \, .\label{eaa1.2}
\end{align} 
It is understood that all quantities
on the right hand side are evaluated at $\vec{t} = \vec{0}$, and we have only displayed those terms on the right hand side in which all the derivatives act on the explicit exponential factor in (\ref{eaa0}). 

We now compare various terms in (\ref{eaa1.2}) with the expected string amplitudes.
The quantity $g_s^{-1} A(\vec{0})$ is the instanton action and  $\exp[{1\over 2}\log g_s + B(\vec 0) ]$
is the exponential of the annulus partition function studied in \cite{Eniceicu:2022nay}.

Ignoring all contributions with vertex operators inserted on closed-string worldsheets, 
since they represent terms where some of the derivatives act on the $\mathcal{Z}^{(0)}$ factor in (\ref{eaa0}),  the leading term for the $n$-point correlation function comes from the product of $n$ disk one-point functions (times the exponentials of the instanton action and the cylinder partition function that accompany all correlation functions).\footnote{In non-critical string theory,  one- and two-point functions on the sphere are non-zero,  so contributions that are more important than this contribution exist. }
This takes the form
\begin{align}
e^{g_s^{-1} A(\vec 0) + {1\over 2}\log g_s + B(\vec 0) } \, g_s^{-n}
\prod_{\alpha=1}^n h_{i_\alpha}\, ,
\label{nptdisk}
\end{align}
where $g_s^{-1} h_{i}$ has the interpretation of the disk one-point function of the vertex operator associated with the $t_i$ deformation.\footnote{We normalize the closed string vertex operators in such a way that they do not carry any factor proportional to $g_s$.}
This matches the leading term in \refb{eaa1.2} if we identify 
\begin{align}
	h_i = \frac{\partial A}{\partial t_i}\, .
	\label{hidadti}
\end{align}
For a given instanton, we choose to pick $t_{i_\alpha}$'s on the left hand side of \refb{eaa1.2} such that the disk one-point function of the operator associated with the $t_{i_\alpha}$ deformation does not vanish.

At the next order,  again ignoring terms with vertex operators inserted on closed-string worldsheets,  we expect three types of contributions:\footnote{So we are not studying contributions from, for instance,  a three-punctured sphere times $(n-3)$ one-punctured disks, and the one-punctured torus times $(n-1)$ one-punctured disks.}
\begin{enumerate}
\item 
    Product of $(n-2)$ disk one-point functions and a disk 
    two-point function. 
    If we denote by
    $g_s f_{ij}$ the ratio of the disk two-point function of operators associated with the $t_i,t_j$
    deformations to the product of the disk one-point functions of the same operators, then this contribution takes the form:
    \be
    e^{g_s^{-1} A(\vec 0) + {1\over 2}\log g_s + B(\vec 0) } \, g_s^{-(n-1)}
    \left\{\prod_{\alpha=1}^n {\p A\over \p t_{i_\alpha}} \right\}  
    \sum_{\beta,\gamma=1\atop \beta<\gamma}^n 
    f_{i_\beta i_\gamma}\, .
    \ee
    \item Product of $(n-1)$ disk one-point functions and an annulus one-point function. 
    If we
    denote by $g_s\, g_i$ the ratio of the annulus one-point function and the disk one-point function
    of the vertex operator associated with the $t_i$ deformation,
    then this contribution takes the form:
    \be
    e^{g_s^{-1} A(\vec 0) + {1\over 2}\log g_s + B(\vec 0) } \, g_s^{-(n-1)}
    \left\{\prod_{\alpha=1}^n {\p A\over \p t_{i_\alpha}} \right\}  
    \sum_{\beta=1}^n 
    g_{i_\beta}\, .
    \ee
    \item Product of $n$ disk one-point functions with the $O(g_s)$ corrections to the instanton action coming from the three-holed sphere and the handle-disk. 
    If we denote by $g_s \wt C$ this $O(g_s)$ correction to the instanton action, this contribution takes the form
    \be
    e^{g_s^{-1} A(\vec 0) + {1\over 2}\log g_s + B(\vec 0) } \, g_s^{-(n-1)}
    \left\{\prod_{\alpha=1}^n {\p A\over \p t_{i_\alpha}} \right\}  
    \wt C\, .\ee
\end{enumerate}
Comparing these three contributions to the three $O(g_s)$ terms inside the square brackets in \refb{eaa1.2}, we arrive at the prediction:
\be\label{eaafgexp}
f_{ij} = { \p^2 A\over \p t_{i} \p {t_{j}}}
\bigg / \left( {\p A\over \p t_{i}} {\p A\over \p t_{j}}
\right) , \qquad g_i= {\p B\over \p t_{i}} \bigg/ {\p A\over \p t_{i}} , \qquad \wt C = C\, .
\ee

So far our analysis involved a general $n$-point function, but now we specialize to the main case of interest in our work.
Consider a special case where all $t_i$'s appearing in \refb{eaa1.2} correspond to deformations of the cosmological constant $\mutilde$ from its background value, which we shall take to be $1$.
The Liouville action (\ref{eq:Liouvilleaction}) tells us that taking a $\mutilde$ derivative brings down an insertion of $- \int \frac{\d x \d y}{\pi} \, e^{2 b \phi}$ on the worldsheet.
Thus, the vertex operator corresponding to the $\mutilde$ deformation is $-V$.

The DDK-KPZ \cite{David:1988hj, Distler:1988jt, Knizhnik:1988ak} scaling, that follows from the shift of the zero mode of the Liouville field in the action (\ref{eq:Liouvilleaction}), implies that the partition function depends on $\mutilde$ and $g_s$ through the combination $g_s^{-1}\mutilde^{Q/(2b)}$. 
Therefore we can take
\begin{align}
g_s^{-1} A(\vec t) &= -g_s^{-1} \mutilde^{Q/(2b)}\, , \label{atmu} \\ 
{1\over 2}\log g_s + B(\vec t) &= 
{1\over 2}\log g_s - {Q\over 4b}\log\mutilde + B_0\, ,
\end{align}
where $B_0$ is a constant that is independent of $\mutilde$ and $g_s$.
As already mentioned,  we normalize $g_s$ so that the instanton action at $\mutilde=1$ equals $-1/g_s$.
Using these results in \refb{eaafgexp} and denoting the $f_{ij}$ and $g_i$ for
$t_i=t_j=\mutilde-1$ by $f$ and $g$ respectively, we get,
\be\label{efgexp}
f = { \p^2 A\over \p \mutilde^2} \bigg / \left({\p A \over \p \mutilde}\right)^2 \bigg \vert_{\mutilde = 1}  
 = {2b\over Q}-1, \qquad 
 g= {\p B \over \p \mutilde} \bigg / {\p A \over \p \mutilde}  \bigg \vert_{\mutilde = 1} =  {1\over 2} \, .
\ee
Our goal will be to verify these relations by explicit worldsheet computations, using string field theory to regularize divergences from the boundaries of moduli space.

We remind the reader that $g_sf$ is the ratio of the disk two-point function to the square of the disk one-point function, and $g_sg$ is the ratio of the annulus one-point function to the disk one-point function.
In particular the mismatch that was observed in \cite{Sen:2020eck} is in the quantity $g$. Here, we have a simple prediction that $g=1/2$, and this will serve as our lamppost to fix the mismatch.
In the end, the direct worldsheet computation, aided by string field theory, leads to perfect agreement with both the predictions $f = 2b/Q - 1$ and $g=1/2$.

As a preliminary sanity check, let us verify that the worldsheet formula (\ref{adiskv}) gives us the disk one-point function of $V = e^{2b\phi}$ that we expect based on the analysis in this section. 
We can use (\ref{hidadti}) and (\ref{atmu}) to get the one-point function of $V$ on the disk, and so we have the prediction
\begin{align}
    A_\text{disk}(V) &= - g_s^{-1} \frac{\partial A}{\partial \mutilde} \bigg\vert_{\mutilde = 1} = 
    g_s^{-1} \, \frac{Q}{2b} \, . 
    \label{adiskv1}
\end{align}
On the other hand, using (\ref{vonepointi}) and (\ref{cghostcorrelator}), the formula (\ref{adiskv}) with $\psi_c = c \cbar V$ yields $ A_\text{disk}(V) = \frac{1}{4g_s} \times \frac{Q}{4b} \times 4 \times 2 = g_s^{-1} \, \frac{Q}{2b}$, as expected.
(The final factor of two is due to the fact that the two terms in (\ref{adiskv}) involving $\partial c$ and $\overline{\partial} \cbar$ give equal contributions.)

\section{The disk two-point function}
\label{sec:disk2pt}
In this section we compute the disk two-point of the cosmological constant operator $V = e^{2 b \phi}$ directly from the worldsheet correlation function and show that we reproduce the value of $f$ given in (\ref{efgexp}).

Using (\ref{fixtwoclosed}) we write the disk two-point amplitude as
\begin{align}
    A_\text{disk}(VV) = \frac{\i}{2g_s} \int_0^1 \d y \, \langle 
    c \overline{c} V(\i) \, (c(z) + \overline{c}(\zbar)) V(z,\zbar)
    \rangle_\text{UHP}\, ,
    \label{adiskvvdef}
\end{align}
where it is understood that the second insertion is at $z = \i y$.
We shall analyze this by dividing the integration region over $y$ into two parts: 
the range $\eps\le y\le 1$
and the range $0\le y\le\eps$ for some small number $\eps$.
We treat the $0\le y\le \eps$ region using string field theory
Feynman diagrams to deal with the potential divergence from the $y\to 0$ region. 
Let us call these two contributions $A_\text{disk}^{(1)}(VV)$ and 
$A_\text{disk}^{(2)}(VV)$, respectively.

First we consider the contribution $A_\text{disk}^{(1)}(VV)$ from the region $y\ge \eps$.
We use the equation of motion (\ref{veom}) and holomorphicity of $c$ to write this as
\begin{align}
    A_\text{disk}^{(1)}(VV) &= \frac{\i}{2bg_s} 
    \int_\epsilon^1 \d y \, \langle c \overline{c} V(\i) \, 
    \left\{\, \overline{\partial} (c\partial \phi(z,\zbar))
    + \partial (\cbar \overline{\partial} \phi(z,\zbar))
    \, \right\}
    \rangle_\text{UHP} \label{avvppbar} \\
    &= \frac{\i}{4bg_s} 
    \int_\epsilon^1 \d y \, \langle c \overline{c} V(\i) \, 
    \left\{\, 
    \partial_x (c \partial \phi(z,\zbar) + \cbar \overline{\partial} \phi(z,\zbar))
    + \i \partial_y (c\partial \phi(z,\zbar) - \cbar\overline{\partial} \phi(z,\zbar))
    \, \right\}
    \rangle_\text{UHP} \,
    \label{avvpxpy}
\end{align}
where we have converted to Cartesian derivatives in the second line.
The term involving $\partial_y$ can be converted to a total derivative, but we need to be a bit more careful with the term involving $\partial_x$.

For this, note that the PSL$(2,\mathbb{R})$ transformation $z \to \frac{z-a}{1+az}$ fixes the puncture at $\i$ but, for small $a$, moves the puncture at $(0,y)$ to $(-a(1-y^2),y)$. That is, it moves the second puncture in the $x$-direction.
If $\p\phi$ and $\bar\p\phi$ had been primaries of dimension $(0,1)$ and
$(1,0)$ respectively, then $c\p\phi$ and $\bar c\bar\p\phi$ would have both been dimension zero primaries, and so PSL$(2,\mathbb{R})$ invariance of the correlation function would imply that the $x$ derivative term in (\ref{avvpxpy}) would vanish.
However, the term proportional to $Q$ in \refb{eftrs1} spoils this argument, since the second derivative with respect to $z$ of $\frac{z-a}{1 + a z}$ does not vanish for $z = \i y$.

We could fix this problem by taking into account the $Q$-dependent terms in (\ref{eftrs1}) for $f(z) = 
\frac{z-a}{1+az}$, but we shall follow a slightly different approach. 
Imagine starting with the disk geometry with the coordinate $w = \frac{1 + \i z}{1 - \i z}$. Also, let $w = r e^{\i\theta}$ in polar coordinates on the disk.
In this geometry, one puncture is located at $w=0$ while the second is placed on the real-$w$ axis at $w = \frac{1-y}{1+y}$.
On the disk geometry, the PSL$(2,\mathbb{R})$ transformation that moves the second puncture off the real axis while keeping the first puncture at the origin fixed is simply a rotation of $w$.
Since this is linear in $w$, the inhomogeneous term in (\ref{eftrs1}) vanishes, and hence the $\theta$ derivative of the correlation function of dimension zero fields vanishes.
So we can first transform (\ref{adiskvvdef}) to the $w$-coordinate system using the fact that $V$ is a dimension $(1,1)$ primary, 
use manipulations analogous to (\ref{avvppbar}) and (\ref{avvpxpy}) to express the result in the $(r,\theta)$ coordinates, drop the $\theta$ derivatives, and finally 
transform it back to the upper half plane using $w = \frac{1 + \i z}{1 - \i z}$.
Using (\ref{eftrs1}), we get
\begin{align}
    c \, \partial\phi(w,\wbar) &= c \, \partial\phi(z,\zbar) + \frac{Q}{z + \i} \, c(z)\,  ,\\
    \cbar \, \overline{\partial}\phi(w,\overline{w}) &= 
    \cbar\, \overline{\partial}\phi(z,\zbar) + \frac{Q}{\zbar - \i}\, \cbar(\zbar)\, .
\end{align}
Plugging these back into the analog of (\ref{avvpxpy}) in the $w$-coordinate system,  we get
\begin{align}
    A_\text{disk}^{(1)}(VV) &= - \frac{1}{4bg_s} \int_\epsilon^1 \d y \, 
    \left\langle c \overline{c} V(\i) \, 
    \partial_y \left(
    c\partial \phi(z,\zbar) + \frac{Q}{z + \i} \, c(z)
    - \cbar \overline{\partial} \phi(z,\zbar)  - \frac{Q}{\zbar - \i}\, \cbar(\zbar)
    \right)
    \right\rangle_\text{UHP} 
    \label{avvtotaldy}
\end{align}

Now we carry out the $y$-integral in $A_\text{disk}^{(1)}(VV)$ by computing the two boundary terms.
The contribution from the $y=1$ boundary is given only by the $c \partial \phi$ and the $\cbar\overline{\partial}\phi$ terms since only these contain the pole necessary to cancel the zero from the OPE of $c(\i) c(z)$.
Thus, using the OPE (\ref{delphiVope}), we get
\begin{align}
    \left. A_\text{disk}^{(1)}(VV)\right\vert_{y=1}
    &= - \frac{1}{4bg_s} \lim_{z \to \i} \, 
    \big \langle 
    c \cbar(\i) \left( c\, \partial\phi(z,\zbar)V(\i) 
    - \cbar \overline{\partial}\phi(z,\zbar) V(\i) 
    \right)
    \big \rangle_\text{UHP} \\
    &= \frac{1}{4g_s} \, \big \langle
    c \cbar \left( \partial c - \overline{\partial}\cbar \right) V
    \big \rangle_\text{UHP} \, ,
\end{align}
where all the insertions in the second line are understood to be at $z = \i$.
We can evaluate this correlator using (\ref{vonepointi}) and (\ref{cghostcorrelator}) to get
\begin{align}
    \left. A_\text{disk}^{(1)}(VV)\right\vert_{y=1}
    &= g_s^{-1} \, \frac{Q}{2b}\, .
    \label{avvc1}
\end{align}

The contribution from the $y = \epsilon$ boundary contains two pieces.
There is a finite piece due to the second and fourth terms in (\ref{avvtotaldy}): 
\begin{align}
    \left. A_\text{disk}^{(1)}(VV)\right\vert_{y=\epsilon,\text{ finite}}
    &= \frac{1}{4bg_s} \cdot \frac{Q}{4b} \cdot \frac{Q}{\i} \cdot (-2\i) \times 2
    = - g_s^{-1} \, \frac{Q^2}{4b^2}\, ,
    \label{avvc2}
\end{align}
where we have again used (\ref{vonepointi}) and (\ref{cghostcorrelator}).
Finally, the contribution from the $y = \epsilon$ boundary contains a divergent piece due to the first and third terms in (\ref{avvtotaldy}).
We use the bulk boundary OPE (\ref{bulkbdryope1}), (\ref{bulkbdryope2}) together with (\ref{vonepointi}) and (\ref{cghostcorrelator}) to get
\begin{align}
    \left. A_\text{disk}^{(1)}(VV)\right\vert_{y=\epsilon,\text{ div}}
    &= \frac{1}{4bg_s} \cdot \frac{Q}{4b} \cdot \left(-\frac{Q}{2\i \epsilon} \right) \cdot (-2\i) \times 2 +O(\epsilon) =  \epsilon^{-1} \,  g_s^{-1} \, \frac{Q^2}{8 b^2} \, .
    \label{avvdiv}
\end{align}
Note that we do not get any order $\epsilon^0$ contribution because of the lack of $O((z - \zbar)^0)$ terms in the bulk-boundary OPE (\ref{bulkbdryope1}), (\ref{bulkbdryope2}).

\begin{figure}[t!]
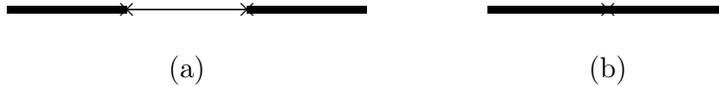

       \hskip7em\hbox{\figthree}
    \vspace{-1.2in}
    \caption{The two Feynman diagrams that contribute to the disk two-point amplitude. The thick lines denote closed strings, while the thin line denotes an open string. 
    The first diagram gives the moduli space integral over the range $0 \leq y \leq \epsilon$,  and the second diagram gives the contribution from the range $\epsilon \leq y \leq 1$.
    Figure reproduced from \cite{Sen:2020eck}.}
    \label{fig:worldsheetdiagramdisk}
\end{figure}
We now analyze the contribution $A_\text{disk}^{(2)}(VV)$ that comes from the region $0 \leq y \leq \epsilon$.
As already mentioned, we shall regard this as the contribution from a Feynman diagram of open-closed string field theory where a pair of open-closed string interaction vertices are joined by an open string propagator. 
This Feynman diagram is shown in figure \ref{fig:worldsheetdiagramdisk}(a).\footnote{The Feynman diagram in figure \ref{fig:worldsheetdiagramdisk}(b) yields the sum of (\ref{avvc1}), (\ref{avvc2}) and (\ref{avvdiv}).}
The two relevant open string states are the tachyon $c_1 \ket{0}$ with $L_0 = -1$, and the out-of-Siegel-gauge mode $\i c_0\ket{0}$ with $L_0 = 0$ \cite{Sen:2020eck}.
Quite generally, in string field theory we expect that the tachyon exchange contribution cancels the divergent piece (\ref{avvdiv}).
We verify this explicitly in appendix \ref{app:cancellation} which also serves as a simple illustration of the mechanism by which string field theory cancels such divergences.
The contribution from the exchange of $c_0\ket{0}$ vanishes since the relevant open-closed vertex, being proportional to the correlator $\langle c \cbar V(\i) \, \partial c(0) \rangle_\text{UHP}$, vanishes (as can be seen from (\ref{cghostcorrelator})).
The contribution from all other states with $L_0 > 0$ vanishes in the limit $\epsilon \to 0$.

Thus, putting together (\ref{avvc1}) and (\ref{avvc2}) we get
\begin{align}
     A_\text{disk}(VV) &= g_s^{-1} \frac{Q^2}{4b^2} \left( \frac{2b}{Q} - 1\right)\, .
\end{align}
Recalling the disk one-point amplitude (\ref{adiskv1}) and the definition of $g_s f$, we get
\begin{align}
    g_s f = \frac{A_\text{disk}(VV)}{A_\text{disk}(V)^2} = g_s \left(
    \frac{2b}{Q} - 1
    \right)\, .
\end{align}
Thus, we find perfect agreement with the prediction (\ref{efgexp}) from the general analysis of section \ref{sec:generalanalysis}.

\section{The annulus one-point function}
\label{sec:annulus1pt}
This section is devoted to analyzing the annulus one-point function of the cosmological constant operator $V$ and deriving the result $g=\frac{1}{2}$ in (\ref{efgexp}) by directly integrating over the moduli space.

We parametrize the annulus with a complex coordinate $w = 2 \pi (x + \i y)$, with the flat metric $\d w \d \wbar$.
As in section \ref{sec:conventions}, we let $2\pi t$ be the Euclidean time in the open string channel, which means that $y$ is periodic with period $t$.
We can fix the translation symmetry along the $y$-axis by setting the $y$-coordinate of the vertex operator to zero.
Thus, we are left with a two-dimensional moduli space, labeled by $v = e^{-2\pi t}$ and $x$.
The range of $v$ is $0 \leq v \leq 1$.
The measure for $v$ integration is proportional to $\d t$ or $\frac{\d v}{v}$, unlike $\frac{\d t}{t}$ for the empty annulus in (\ref{eq:emptyannulus}), since the translation symmetry in the $y$-direction has been fixed.
The open string has length $\pi$, so $x$ spans the range $0 \leq x \leq \frac{1}{2}$.
However, note that $w \to \pi - w$ is a diffeomorphism of the annulus, since both boundaries lie on the same instanton.
We need to quotient by diffeomorphisms, and so the range of integration of $x$ is in fact $0 \leq x \leq \frac{1}{4}$.

Recall that $g_s g$ is defined as the ratio of the annulus one-point amplitude to the disk one-point amplitude.
By general principles, we have\footnote{
See, for example, section 7.3 of \cite{Polchinski:1998rq} for the analogous statement for the torus amplitude with external closed string states.} 
\begin{align}
    g_s\, g &= \int_0^1 \d v \int_{0}^\frac{1}{4} \d x \, F(v,x)
    = g_s \, \frac{4b}{Q} \int_0^1 \d v \int_{0}^\frac{1}{4} \d x  \,  \Tr\left[V(w,\bar w) \, b_0 \, c_0
\, v^{L_0-1}\right]\, . \label{g_formula_1}
\end{align}
The nontrivial proportionality constant in the final expression is determined in appendix \ref{sec:annulus_overall_constant}. The coordinate $w$ should not be confused with the
local coordinates around the punctures, to be discussed in section \ref{sec:vertices_review},
for which we use the same symbol.

\begin{figure}[t!]
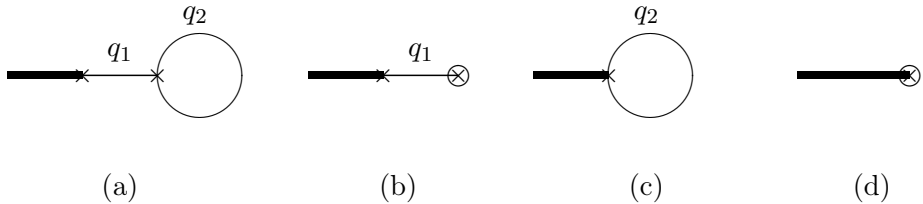

    \begin{center}
        \hbox{\figsix}
    \end{center}
    \vspace{-1.6in}
    \caption{The four Feynman diagrams that contribute to the annulus one-point amplitude. The thick line denotes a closed string, while thin lines denote open strings. 
    The $\times$ denotes a vertex on the upper half plane, whereas $\otimes$ denotes a vertex on the annulus.
    The variables $q_1$ and $q_2$ are the plumbing fixture variables associated with the corresponding propagators.
    Figure reproduced from \cite{Sen:2020eck}.}
    \label{fig:worldsheetdiagrams}
\end{figure}

The integral (\ref{g_formula_1}) has divergences for small $v$ and small $x$.
As explained in detail in \cite{Sen:2020eck}, we need to interpret the contributions from these regions as string field theory Feynman diagrams with internal propagators.
These Feynman diagrams are shown in figure \ref{fig:worldsheetdiagrams} and the corresponding regions in the moduli space are shown in figure \ref{fig:vxmodulispace}.
The Feynman diagram (d) represents the contributions from the ``bulk'' of moduli space where there is no degeneration.
Feynman diagram (c) has one internal propagator and corresponds to small $v$ with finite $x$.
Feynman diagram (b) also has one internal propagator, but corresponds to small $x$ with finite $v$.
Finally, Feynman diagram (a) has two internal propagators and corresponds to small $v$ \emph{and} small $x$. 

Before diving into the computational details, we give an overview of the construction of the various vertices that we will need.
We will be brief since all the details were explained in \cite{Sen:2020eck}.

\subsection{Brief review of the construction of vertices}
\label{sec:vertices_review}
We will discuss the five different types of vertices that appear in figure \ref{fig:worldsheetdiagrams}: (1) the upper half plane C-O vertex, (2) the upper half plane O-O-O vertex, (3) the upper half plane C-O-O vertex, (4) the annulus O vertex, and (5) the annulus C vertex. Here C and O stand respectively for closed and open strings.

The C-O vertex is described by the upper half plane geometry with complex coordinate $z$. 
The closed string puncture is
inserted at $z = \i$ and the open string puncture is inserted at $z =0$.
There are no moduli.
The local coordinate $w$ around the open string puncture is taken to be $w = \lambda z$, with $\lambda$ being a large, fixed real number.

The O-O-O vertex is described by the upper half plane geometry. 
Let's again denote the complex coordinate on the upper half plane by $z$, with the three open string punctures inserted at $z=0$, $z=1$ and $z = \infty$.
There are no moduli, but in computing a physical amplitude, one needs to sum over the two distinct cyclic permutations of the three points.
Let $\alpha$ be another large, fixed real parameter.  
The local coordinates around the three punctures are chosen to be
\begin{align}
    w_1 = \alpha\, \frac{2z}{2-z} \, , \quad 
    w_2 = \alpha \, \frac{2(z-1)}{z+1} \, , \quad 
    w_3 = \alpha \, \frac{2}{1-2z}\, .
    \label{ooo-w}
\end{align}

Next, we discuss the C-O-O vertex, which appears in figure \ref{fig:worldsheetdiagrams}(c).
The C-O-O amplitude on the upper half plane has a one-dimensional moduli space, and provides an example of how the boundaries of moduli space are assigned to Feynman diagrams with internal propagators.
Denoting the complex coordinate on the upper half plane by $z$, we insert the closed string puncture at $z = \i$ and the two open string punctures at $z = \pm \beta$.
Here $\beta$ is a positive real number that can be taken to be the coordinate on the one-dimensional moduli space.
The PSL$(2,\mathbb{R})$ transformation $z \to -1/z$ keeps the puncture at $z = \i$ fixed, but sends $\beta \to -1/\beta$.
So we can restrict $\beta$ to lie in the range $0 \leq \beta \leq 1$, while summing over the two permutations of the two open string punctures.
The region of small $\beta$, when the two open string punctures collide, is covered by the Feynman diagram that consists of an upper half plane C-O vertex (with coordinate $z$) joined to an upper half plane O-O-O vertex (with coordinate $\zt$) with an open string propagator.
The joining happens via the plumbing fixture relation $\lambda z \cdot \alpha \frac{2\zt}{2-\zt} = - q_1$, with $0 \leq q_1 \leq 1$.
Now define
\begin{align}
    \lt := \lambda \alpha    \, .
\end{align}
The open string puncture at $\zt = 1$ gets mapped to $z = - \frac{q_1}{2\lt}$, while the one at $\zt = \infty$ gets mapped to $z = + \frac{q_1}{2\lt}$. 
Thus, we see that $\beta = \frac{q_1}{2\lt}$.
Since the range of $q_1$ is $0 \leq q_1 \leq 1$, we see that this Feynman diagram covers the region $0 \leq \beta \leq \frac{1}{2\lt}$.
Note that $\lt$ is large, so this is the region of small $\beta$, where the two open-string punctures are close to each other.
The upper half plane C-O-O vertex is then assigned to cover the remaining region $\frac{1}{2\lt} \leq \beta \leq 1$.
For the range $0 \leq \beta \leq \frac{1}{2\lt}$, the local coordinates around the two open string punctures are induced by the choice of local coordinates in the C-O and O-O-O vertices described above.
For the range $\frac{1}{2\lt} \leq \beta \leq 1$, we need to pick a choice of local coordinates that, at $\beta = \frac{1}{2\lt}$, match the ones from the $\beta\leq \frac{1}{2\lt}$ region.
The choice made in equation (4.12) of \cite{Sen:2020eck} involved a real-valued function $f(\beta)$ that is only constrained by the following 
\begin{align}
    f\left(\frac{1}{2\lt} \right) = \frac{4 \lt^2 - 3}{8 \lt^2} \, , \quad f(1) = 0 \, , \quad f(-\beta) = -f(\beta)\, .
    \label{fvalues}
\end{align}
From now on,  we shall work in the limit of large $\alpha$ and $\wt\lambda$ and ignore terms
that are suppressed by inverse powers of either $\alpha$ or $\wt\lambda$. 
Of course, the final result is guaranteed to be independent of $\alpha$, $\wt\lambda$ and the function $f(\beta)$.

Next, we discuss the vertex that corresponds to one open string puncture on the annulus. 
This appears as one of the vertices in figure \ref{fig:worldsheetdiagrams}(b).
There is one modulus, the quantity $v$ defined in (\ref{vdef}).
The full range of $v$ is $0 \leq v \leq 1$.
As discussed in \cite{Sen:2020eck}, the region 
$0 \leq v \leq (\alpha^2 - \frac{1}{2})^{-1}$ of small $v$ corresponds to the upper half plane O-O-O vertex with two of the three open string punctures joined with a propagator.
The remaining range $(\alpha^2 - \frac{1}{2})^{-1} \leq v \leq 1$ is assigned to the annulus O vertex.
The choice of the local coordinate around the open string puncture has been described in \cite{Sen:2020eck}.

\begin{figure}[t!]
    \centering
    \includegraphics[width=0.5\textwidth]{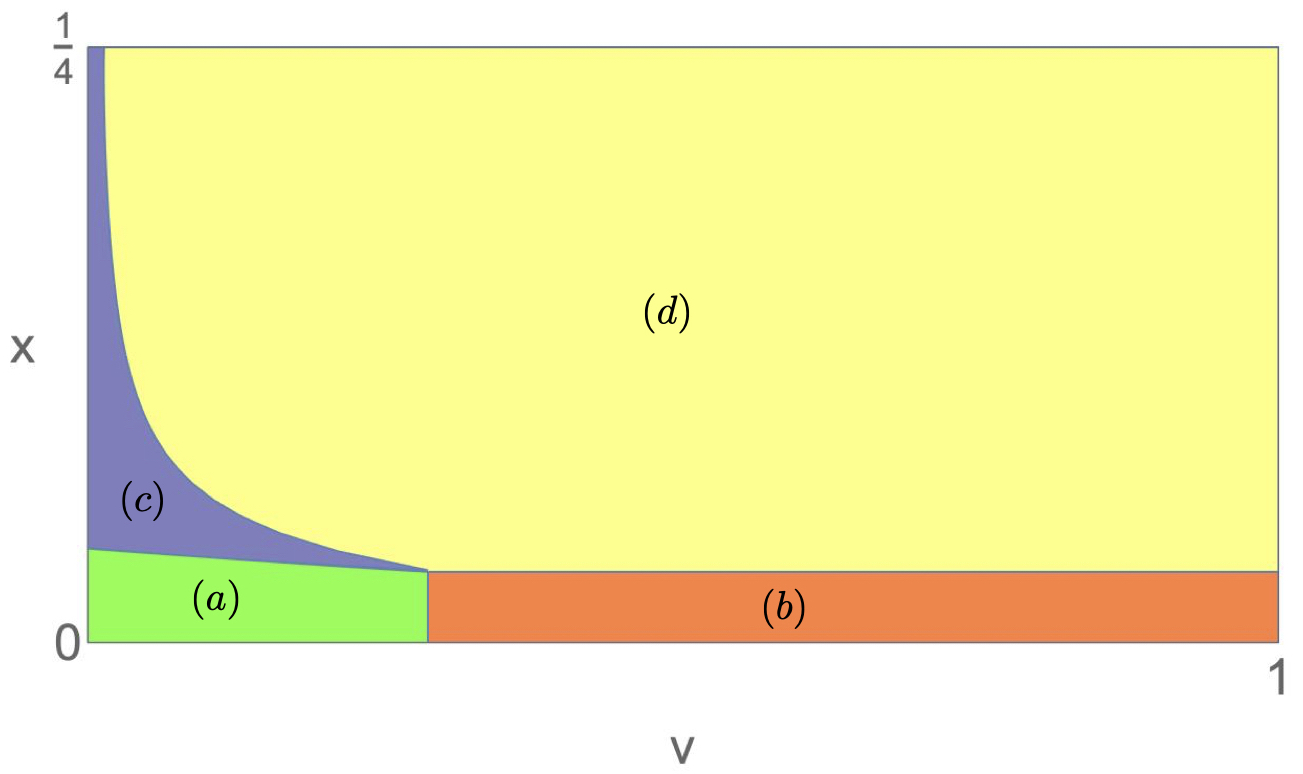}
    \caption{The division of moduli space of the annulus with one bulk puncture into four regions corresponding to the four Feynman diagrams in figure \ref{fig:worldsheetdiagrams}. 
    The green region describes the Feynman diagram in figure \ref{fig:worldsheetdiagrams}(a) and is given in (\ref{aregion}). 
    The red region describes the Feynman diagram in figure \ref{fig:worldsheetdiagrams}(b) and is given in (\ref{bregion}).
    The blue region describes the Feynman diagram in figure \ref{fig:worldsheetdiagrams}(c) and is given in (\ref{cregion1})-(\ref{cregion3}).
    The remaining yellow region describes the Feynman diagram in figure \ref{fig:worldsheetdiagrams}(d), and covers the bulk of the $(v,x)$ moduli space.
    We have taken $\lt = 4$, $\alpha = 2$ and $f(\beta) = \frac{4\lt^3-3\lt}{4\lt^2-1} \beta(1-\beta^2)$ for the purposes of plotting this figure.
    }
    \label{fig:vxmodulispace}
\end{figure}

Finally, we come to the C amplitude on the annulus.
As already explained, in this case there is a two-dimensional moduli space parametrized by $v$ and $x$.
We need to divide this moduli space into four regions \cite{Sen:2020eck}, which correspond to the four Feynman diagrams in figure \ref{fig:worldsheetdiagrams}.
See also figure \ref{fig:vxmodulispace}.
The Feynman diagram in figure \ref{fig:worldsheetdiagrams}(a) corresponds to the region (shown in green in figure \ref{fig:vxmodulispace})
\begin{align}
    0 \leq v \leq \left(\alpha^2 - \frac{1}{2}\right)^{-1}\, , \quad 
    0 \leq 2\pi x \leq \lt^{-1} \, \frac{2-v}{2+v}\, ,
    \label{aregion}
\end{align}
with $(v,x)$ related to the plumbing fixture variables $(q_1, q_2)$ as
\begin{align}
    v = \frac{q_2}{\alpha^2} \left(1 - \frac{q_2}{2\alpha^2} \right)^{-1} \, , \quad 
    2\pi x = \frac{q_1}{\lt} \left(1 - \frac{q_2}{\alpha^2} \right)\, .
    \label{aregion_vx_q1q2}
\end{align}
The plumbing fixture variables always vary in the range $[0,1]$.
The Feynman diagram in figure \ref{fig:worldsheetdiagrams}(b) corresponds to the region (shown in red in figure \ref{fig:vxmodulispace})
\begin{align}
    \left(\alpha^2 - \frac{1}{2} \right)^{-1} \leq v \leq 1 \, , \quad 
    0 \leq 2\pi x \leq \lt^{-1} (1 - \alpha^{-2})\, ,
    \label{bregion}
\end{align}
with $x$ related to the plumbing fixture variable $q_1$ as
\begin{align}
    2\pi x = \frac{q_1}{\lt} \left(1 - \frac{1}{\alpha^2} \right)\, .
    \label{xq1_2b}
\end{align}
The Feynman diagram in figure \ref{fig:worldsheetdiagrams}(c) corresponds to the region in the $(v,x)$ plane parametrized as 
\begin{align}
    \frac{1}{2\lt} \leq \beta \leq 1 \, ,& \quad 
    0 \leq u \leq \alpha^{-2} \left( 1 + \frac{1}{4\lt^2} \right)^{-2}\, , 
    \quad \text{with} \label{cregion1} \\
    2 \pi x(\beta, u) &= 2 \tan^{-1}(\beta) 
    - \frac{u}{\beta \lt^2} \left( 1 - \beta^2 - 2 \beta f(\beta) \lt \right)
    \label{cregion2} \quad \text{and } \\
    v(\beta, u) &= u \frac{(1+\beta^2)^2}{4\beta^2 \lt^2} \left( 
    1 + \frac{u}{2\beta^2 \lt^2} 
    \left(1 - \beta^2 - 2 \beta f(\beta)\lt \right)^2
    \right) \, . \label{cregion3}
\end{align}
The parameter $u$ is related to the plumbing fixture variable $q_2$ via 
$u = q_2 \alpha^{-2}(1+\frac{1}{4\lt^2})^{-2}$.
Note that this corresponds to small $v$ but finite $x$ region, 
and is shown in blue in figure \ref{fig:vxmodulispace}.
The Feynman diagram in figure \ref{fig:worldsheetdiagrams}(d) covers the remaining $(v,x)$ region, not included in the three cases above;
it is shown in yellow in figure \ref{fig:vxmodulispace}.

We will need to integrate a total derivative on region (d), and so we need to discuss the boundaries of region (d).
There are four boundary components, see figure \ref{fig:vxmodulispace}.
The boundary between regions (d) and (b) lies at fixed $x = (2\pi\lt)^{-1}(1-\alpha^{-2}) $ and is parametrized by $v$ in the range given in (\ref{bregion}).
The boundary between regions (d) and (c) is parametrized by $\beta$ in the range given in (\ref{cregion1}), with fixed $u = \alpha^{-2}(1 + \frac{1}{4\lt^2})^{-2}$ and $x,v$ given by
\refb{cregion2}, \refb{cregion3}.
The top boundary lies at $x=\frac{1}{4}$, with $ (\lt \alpha)^{-2} ( 1 + \frac{1}{4\lt^2} )^{-2} \leq v\leq 1$.
Finally, the right boundary lies at $v=1$ with $ (2\pi\lt)^{-1}(1-\alpha^{-2})  \leq x \leq \frac{1}{4}$.

We end this subsection with a couple of important remarks. 
First, in doing the computations, we will be expanding various expressions for large $\alpha$ and large $\lt$. 
In doing so, we can drop all terms that contain negative powers of either variable $\alpha$ or $\lt$, since these cannot contribute to the final answer \cite{Sen:2020eck}.
The reason is that the $\alpha$ and $\lt$ dependent terms are supposed to cancel in the end, and if a term contains a negative power of one large variable, there is no way for it to give a $\alpha$ and $\lt$ independent term at higher orders.
Second, for diagrams that contain an internal propagator, we need the string field theory replacement rules \cite{Sen:2020eck}
\begin{align}
    \int_0^{1} \d q \, q^{-2} &\to -1 \, , \quad 
    \int_0^{1} \d q \, q^{-1} \to 0 \, .
    \label{replacementrule}
\end{align}
Here $q$ is the parameter that enters in the plumbing fixture relation.
The first replacement rule comes from the formal Schwinger parameter representation of the propagator of a field with $L_0 = -1$.
The second replacement rule comes from the fact that the Siegel gauge zero modes are not part of the SFT path integral.
They are replaced by the $\psi$ field that multiples $\i c_0\ket{0}$, and the contribution of $\psi$ exchanges is taken into account separately in section \ref{sec:psiexchange}.

\subsection{The worldsheet contribution}
\label{sec:gws}

In this subsection, we will compute the worldsheet contribution to $g$.
There are two other contributions, which we shall compute in the subsequent subsections.

Let us begin by computing the contribution from the ``bulk'' region of moduli space, corresponding to the Feynman diagram in figure \ref{fig:worldsheetdiagrams}(d). 
Let us denote the region of the $(v,x)$-plane covered by this diagram by $S$.
We choose $S$ to have the orientation given by the two-form $\d v \wedge \d x$.
We use the equation of motion (\ref{veom}) with $\mutilde = 1$ to replace $V$ in (\ref{g_formula_1}) by $\frac{1}{b} \partial_w \partial_{\wbar} \phi$, or $\frac{1}{16 \pi^2 b} (\partial_x^2  + \partial_y^2)\phi$.
The $\partial_y^2 \phi$ term does not contribute, since the transformation that moves the vertex operator insertion in the $y$-direction is a simple translation for which the anomaly term in the transformation of $\partial \phi$ (\ref{eftrs1}) vanishes.
Thus
\begin{align}
    g^{(d)} = \frac{1}{4\pi^2 Q} \int_S \d v \, \d x\, 
    \partial_x 
    \Tr\left[ \partial_x \phi(w, \wbar) \, b_0 \, c_0\, v^{L_0-1}\right]
    = g_s^{-1} \int dv \, dx \, \p_x G(v,x)\, ,
    \label{gd_def_dx2phi}
\end{align}
where $G(v,x)$ has been defined in \refb{c3}, \refb{cresultannulus}.
Since the right hand side of this equation is a total derivative, the computation reduces to integrating $\Tr\left[ \partial_x \phi(w, \wbar) \, b_0 \, c_0\, v^{L_0-1}\right]$ along the boundary of $S$.

The boundary of $S$ has four components, as discussed at the end of section \ref{sec:vertices_review}.
The top boundary at $x=\frac{1}{4}$ does not contribute since the symmetry $w \to \pi - w$ implies that $\langle \partial_x\phi(w=\frac{\pi}{2}) \rangle_\text{annulus} = 0$.
There are no divergences at the $v=1$ boundary, and since we are integrating $\partial_x (\ldots)$ on $S$, via Stokes's theorem, this boundary also does not contribute.
So we only need to consider the two remaining boundaries: the boundary between regions (d) and (c), and the boundary between regions (d) and (b).

Let us first consider the boundary between regions (d) and (b), and denote the contribution as $g^{(b)\mhyphen(d)}$.
This boundary lies at constant small $x$, namely $x = \frac{1-\alpha^{-2}}{2\pi\widetilde{\lambda}}$, with $v$ lying in the range given in the first part of (\ref{bregion}).
The small-$x$, finite-$v$ behavior can be found from \refb{eGexp} and yields,
\begin{align}
    g^{(b)\mhyphen(d)} = 
    \frac{1}{2\pi} \, \widetilde{\lambda}\,  (1-\alpha^{-2})^{-1}
    \int_{(\alpha^2 - \frac{1}{2})^{-1}}^1 \d v \, \frac{Z(v)}{v}\, .
\end{align}
We could evaluate this using the known form of $Z(v)$, but we note that this contribution will be exactly cancelled by the contribution $g^{(b)}$ of the Feynman diagram in figure \ref{fig:worldsheetdiagrams}(b).\footnote{This is analogous to a similar argument made in appendix D of \cite{Sen:2020eck}.}
Physically, this happens because there are no $L_0 = 1$ states in the Liouville open string sector that can leave an order-one contribution, and because the relation between $x$ and $q_1$ in (\ref{xq1_2b}) is linear.
More precisely, to evaluate the Feynman diagram in figure \ref{fig:worldsheetdiagrams}(b), we use the second line of (\ref{efvxtwonew}), write $x$ in terms of $q_1$ using (\ref{xq1_2b}), and use the replacement rule $\int_0^1 \frac{\d q_1}{q_1^2} \to -1$.

Now let us consider the boundary between regions (d) and (c), and denote the contribution as $g^{(c)\mhyphen(d)}$.
As discussed at the end of section \ref{sec:vertices_review}, the boundary between regions (d) and (c) is parametrized by $(v(\beta,u), x(\beta,u))$ given in (\ref{cregion2}) and (\ref{cregion3}), with $\frac{1}{2\lt} \leq \beta \leq 1$ and $u$ fixed to be $u = \alpha^{-2}(1 + \frac{1}{4\lt^2})^{-2}$.
Let us denote (\ref{cregion2}) and (\ref{cregion3}) with this fixed value of $u$ by $(v(\beta), x(\beta))$.
Since $v$ is small along this boundary, we need to understand the small-$v$ behavior of 
$G(v,x)$.
This has been given in (\ref{eGexp}) of appendix \ref{sec:annulus_overall_constant}, using which we get
\begin{align}
  g^{(c)\mhyphen(d)}  = -\frac{1}{2\pi} \int \d v \, (v^{-2} - 2 v^{-1}) \cot (2\pi x)\, ,
  \label{gcd}
\end{align}
with the direction of integration being the direction of increasing $\beta$, that is, upwards along the boundary of the blue and yellow regions in figure \ref{fig:vxmodulispace}.
To do the integral in (\ref{gcd}), it is helpful to first rewrite it as
\begin{align}
    g^{(c)\mhyphen(d)} = \frac{1}{2\pi} \int_{(2\widetilde{\lambda})^{-1}}^1 \d \beta \, \frac{\partial}{\partial\beta} \left( 
    v(\beta)^{-1} + 2 \log v(\beta)
    \right) \cot (2\pi x(\beta)) \, .
\end{align}
Then carrying out the integrations using the known expressions for $v(\beta)$ and $x(\beta)$ given above, we get
\begin{align}
    g^{(c)\mhyphen(d)} = 
    \frac{\alpha^2 \widetilde{\lambda}^2}{4} - \frac{\alpha^2 \widetilde{\lambda}}{\pi}
    - \frac{3\widetilde{\lambda}}{4\pi} + \frac{\alpha^2}{8} 
    - \frac{2\widetilde{\lambda}^2}{\pi} \int_{(2\widetilde{\lambda})^{-1}}^1 \d \beta \, \frac{f(\beta)^2}{1+\beta^2}
    + \frac{1}{2} 
    \, .
    \label{gcdfinal}
\end{align}

Now, let us evaluate the contribution from the Feynman diagram in figure \ref{fig:worldsheetdiagrams}(c).
This diagram corresponds to small $v$ but finite $x$, the precise domain has been specified in (\ref{cregion1})-(\ref{cregion3}).
Using (\ref{cregion2}) and (\ref{cregion3}) we get the transformation of the measure
\begin{equation}
     \d\qt \,\d x  = \left( \frac{1 + \beta^2}{4\pi \beta^2 \lt^2} + O(u) \right) \d \beta \, \d u,
\end{equation}
and so, using (\ref{g_formula_1}) and the first line of (\ref{efvxtwonew}), we have
\begin{align}
    g^{(c)} 
    &= \int_{(2\lt)^{-1}}^{1} \d\beta \int_{0}^{\alpha^{-2}(1 + (4 \lt^2)^{-1})^{-2}} \d u\, 
    \left( \frac{1 + \beta^2}{4\pi \beta^2 \lt^2}  +O(u) \right)
    \frac{1}{\sin^2(2\pi x)} 
    \left( v^{-2} - 2 v^{-1} + O(v^0) \right) \nonumber \\
    & \to - \alpha^{2} \left(1 + \frac{1}{4 \lt^2} \right)^2 \int_{(2\lt)^{-1}}^{1} \d\beta \, \frac{\lt^2}{\pi(1+\beta^2)} 
    \approx 
    - \frac{\alpha^2 \lt^2}{4} + \frac{\alpha^2 \lt}{2\pi} - \frac{\alpha^2}{8}\, .
    \label{gc}
\end{align}
In the second step, we used (\ref{cregion2}) and (\ref{cregion3}), 
replaced $u$ by $q_2 \alpha^{-2}(1+\frac{1}{4\lt^2})^{-2}$, applied the replacement rule (\ref{replacementrule}) for the plumbing fixture variable $q_2$, simplified a bit and ignored
terms containing inverse powers of either $\alpha$ or $\lt$. 
The $O(u)$ term from the measure can combine with the $v^{-2}$ term to give an $O(1)$ expression proportional to $\int du/u=\int dq_2/q_2$, but this is set to zero using the replacement rule  \refb{replacementrule}.

The contribution from the Feynman diagram in figure \ref{fig:worldsheetdiagrams}(a) can be obtained by using the expression for $F(v,x)$ in (\ref{efvxtwonew}) for small $v$ and small $x$, rewriting the answer in terms of the plumbing fixture variables $q_1$ and $q_2$ using (\ref{aregion_vx_q1q2}), and then using the first replacement rule in (\ref{replacementrule}) for both $q_1$ and $q_2$.
We get
\begin{align}
    g^{(a)} = \frac{1}{2\pi} \lt \, \alpha^2 \, .
    \label{ga}
\end{align}

Putting together all the Feynman diagrams in figure \ref{fig:worldsheetdiagrams} using equations (\ref{ga}), (\ref{gc}), (\ref{gcdfinal}), and the fact that $g^{(b)} + g^{(b)\mhyphen(d)} = 0$, we get the total worldsheet contribution to $g$ as
\begin{align}
    g_\text{ws} &= g^{(a)}+g^{(b)}+g^{(c)}+ \left( g^{(b)\mhyphen(d)} + g^{(c)\mhyphen(d)} \right) \\
    &= \frac{1}{2} - \frac{3\lt}{4\pi} - \frac{2 \lt^2}{\pi} \int_{(2\widetilde{\lambda})^{-1}}^1 \d \beta \, \frac{f(\beta)^2}{1+\beta^2} \, .
    \label{gws}
\end{align}

\subsection{Contribution from $\psi$ exchange}
\label{sec:psiexchange}

\begin{figure}[t!]
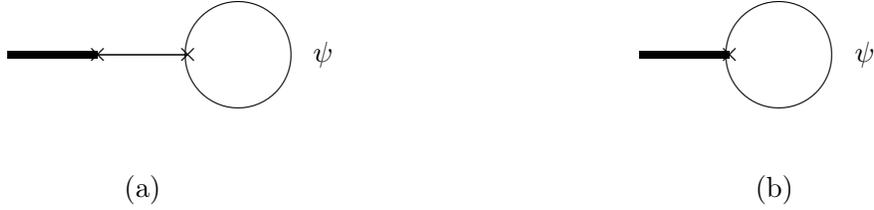

    \begin{center}
        \hbox{\figpsione}
    \end{center}
    \vspace{-2.0in}
    \caption{The two Feynman diagrams involving the loop of the out-of-Siegel-gauge mode $\psi$. They both involve the upper half plane C-O-O amplitude; (a) corresponds to small $\beta$ while (b) corresponds to finite $\beta$. Figure adapted from \cite{Sen:2020eck}.}
    \label{fig:worldsheetdiagrampsi}
\end{figure}

Since the two Siegel-gauge zero modes get replaced by the out-of-Siegel gauge mode $\psi$ that multiplies the state $\i c_0\ket{0} = \i \partial c(0)\ket{0}$,  we need to take into account the contribution from the diagrams involving 
$\psi$ propagators explicitly \cite{Sen:2020eck}.
This is a contribution that is not captured by the string worldsheet.

As mentioned in section \ref{sec:disk2pt}, the one-point C-O disk amplitude with one on-shell closed string and the open string puncture corresponding to $\psi$ vanishes.
So, we only need to consider the two Feynman diagrams shown in figure \ref{fig:worldsheetdiagrampsi}; they are analogous to figure \ref{fig:worldsheetdiagrams}(a) and \ref{fig:worldsheetdiagrams}(c).
Both diagrams involve an upper-half-plane amplitude with one closed string puncture, and two open string punctures joined with a $\psi$ propagator.
Following the discussion of the upper half plane C-O-O amplitude in section \ref{sec:vertices_review}, we see that the region $0 \leq \beta \leq \frac{1}{2\lt}$ corresponds to figure \ref{fig:worldsheetdiagrampsi}(a), while the region 
$ \frac{1}{2\lt} \leq \beta \leq 1$ corresponds to figure \ref{fig:worldsheetdiagrampsi}(b).
As discussed in section \ref{sec:ghostconventions}, the $\psi$ propagator equals $\frac{1}{2}$ in our conventions.

The geometry of the upper half plane C-O-O amplitude has been discussed in section \ref{sec:vertices_review}.
We represent the upper half plane with complex coordinate $z$, insert the closed string puncture at $z = \i$ and insert the two open string punctures at $z_{1} = -\beta$ and $z_2 = \beta$.
Let $w_a$, with $a \in \{1,2\}$ denote the local coordinates around the two open string punctures, which we relate to $z$ as
\begin{align}
    z  = F_a(w_a, \beta) = z_a + g_a(\beta) w_a + \frac{1}{2} h_a(\beta) w_a^2 + O(w_a^3)\, .
\end{align}
The actual amplitude, normalized to directly give the contribution to $g$, is given by \cite{Zwiebach:1992ie, Zwiebach:1997fe, deLacroix:2017lif} 
\begin{align}
    g_\psi = - K_1 \int \d \beta \sum_{a=1}^2 \oint_a \frac{\d z}{2\pi \i} 
    \frac{\partial F_a}{\partial \beta} \, \langle 
    b(z) F_1 \circ \i\partial c(0) \, F_2 \circ \i\partial c(0) \, c\cbar V(\i)
    \rangle_\text{UHP}\, ,
    \label{gpsidef}
\end{align}
where $\oint_a$ represents an anticlockwise contour around $z_a$, and $F_a \circ \i \partial c(0)$ denotes the conformal transformation of the operator $\i\partial c$ by the function $F_a$.
The constant $K_1$ is a normalization constant that we determine in appendix \ref{app:psinormalization} by comparing the right hand side of (\ref{gpsidef}), with $\psi$
replaced by the tachyon,  to the result for either figure 
\ref{fig:worldsheetdiagrams}(c) or \ref{fig:worldsheetdiagrams}(a), which were given in (\ref{gc}), \refb{ga}.
The result is $K_1 = \frac{\i b}{2\pi Q}$. 
This includes the contribution of the $\psi$ propagator. 
The $\beta$ integral captures the effect of the Schwinger parameter integral associated with the horizontal open string propagator in figure \ref{fig:worldsheetdiagrampsi}(a), and the intrinsic integration parameter of the C-O-O vertex in figure \ref{fig:worldsheetdiagrampsi}(b).

Explicitly, the local coordinates around the open string punctures are \cite{Sen:2020eck}
\begin{align}
    F_1(w_1, \beta) &= - \beta + \frac{2\beta}{\alpha} w_1 - \frac{\beta}{\alpha^2} w_1^2 + O(w_1^3) \\
    F_2(w_2, \beta) &=  \beta + \frac{2\beta}{\alpha} w_2 + \frac{\beta}{\alpha^2} w_2^2 + O(w_2^3)
\end{align}
for $0 \leq \beta \leq \frac{1}{2\lt}$, and 
\begin{align}
    F_1(w_1, \beta) &= 
    - \beta 
    + \frac{4\lt(1+\beta^2)}{\alpha(4 \lt^2+1)} w_1 
    - \frac{16 \lt^2 (1+\beta^2)(\beta + \lt f(\beta))}{\alpha^2(4\lt^2 + 1)^2} w_1^2 + O(w_1^3) \label{f1wbeta} \\
     F_2(w_2, \beta) &= 
     \beta 
    + \frac{4\lt(1+\beta^2)}{\alpha(4 \lt^2+1)} w_2 
    + \frac{16 \lt^2 (1+\beta^2)(\beta + \lt f(\beta))}{\alpha^2(4\lt^2 + 1)^2} w_2^2 + O(w_2^3) \label{f2wbeta}
\end{align}
for $\frac{1}{2\lt} \leq \beta \leq 1$.

The evaluation of $g_\psi$ given in (\ref{gpsidef})  involves simplifying the right hand side
using the $bc$ OPE, the explicit form of the local coordinates given above, and using (\ref{cghostcorrelator}) and (\ref{vonepointi}) to calculate the correlator. 
Using $K_1 = \frac{\i b}{2\pi Q}$, we find
\begin{align}
    g_\psi = \frac{\lt}{4\pi} + 
    \frac{2 \lt^2}{\pi} 
    \int_{(2\widetilde{\lambda})^{-1}}^1 \d \beta \, \frac{f(\beta)^2}{1+\beta^2}\, .
    \label{gpsi}
\end{align}
The first term is the contribution from the Feynman diagram in figure \ref{fig:worldsheetdiagrampsi}(a), 
and the second term is the contribution from the Feynman diagram in figure \ref{fig:worldsheetdiagrampsi}(b). 

\subsection{Gauge parameter redefinition}
\label{sec:gghost}

Finally, we also need to take into account the relation between the string field theory gauge transformation parameter $\theta$ appearing in (\ref{pq_to_psi}) and the rigid $U(1)$ transformation parameter $\widetilde{\theta}$ under which an open string with one end on the instanton picks up a phase $e^{\i\widetilde{\theta}}$.
The gauge transformation parameter $\theta$ is the one that multiplies the vacuum state 
$\i \ket{0}$.
Once this relation is found, we can evaluate the denominator of (\ref{pq_to_psi}) using the fact that $\widetilde{\theta}$ has period $2\pi$.

In order to study the gauge transformation properties, it is useful to introduce a spectator instanton.
A fundamental property of the BV formalism is that the gauge transformation laws are encoded in the coupling terms in the master action.
Let $\xi$ be the field that multiplies the vacuum $\ket{0}$ in the expansion of the string field with Chan-Paton factor $\begin{pmatrix} 0 & 1 \\ 0 & 0 \end{pmatrix}$.
Its conjugate anti-field $\xi^*$ has the vertex operator $c \partial c \partial^2 c/2$ and carries the Chan-Paton factor $\begin{pmatrix} 0 & 0 \\ 1 & 0 \end{pmatrix}$.
Let $\psi_2$ be the field that multiplies the vacuum $i\ket{0}$ but with Chan-Paton factor $\begin{pmatrix} 1 & 0 \\ 0 & 0 \end{pmatrix}$.
Computing the trilinear coupling of $\xi$-$\xi^\star$-$\psi_2$ using the correlator (\ref{cghostcorrelator}), we find the following term in the SFT path integral \cite{Sen:2020eck, Sen:2021qdk}:
\begin{align}
    \exp(\i\, g_o \, \xi \xi^\star \psi_2)\, .
\end{align}
This coupling determines the infinitesimal gauge transformation of $\xi$ to be
$\delta\xi=\i g_o\theta\xi$ and, in turn, gives the
relationship between $\theta$ and $\widetilde{\theta}$ to be $\d\theta = g_o^{-1}\d\widetilde{\theta}$. This is needed for a precise computation of the one-loop normalization factor (the exponential of the empty annulus) that accompanies D-instanton amplitudes \cite{Sen:2021qdk}.

However, we are working at the first sub-leading order in $g_s$, and the relation between $\theta$ and $\widetilde{\theta}$ could receive corrections that are, in general, field-dependent.
Let $\Phi$ be the tachyon field that multiplies the cosmological constant operator $V$ in the expansion of the string field, and let $\i\mathcal{A}$ be the C-O-O-O disk amplitude with the closed string insertion being $\Phi$ and the three open string insertions being $\xi$, $\xi^\star$ and $\psi_2$.
Following \cite{Sen:2020eck}, we note that this leads to the following field-dependent relationship between $\theta$ and $\widetilde{\theta}$:
\begin{align}
     \d\wt\theta = g_o\left(1 + \mathcal{A} \Phi \right)  \d{\theta}\, .
\end{align}
Substituting this in the denominator of (\ref{pq_to_psi}), and writing $(1 + \mathcal{A} \Phi) \approx e^{\mathcal{A}\Phi}$, we see that the path integral contains the extra term $\exp(\mathcal{A}\Phi)$.
This leads to an additional contribution to the one-point function of $\Phi$ proportional to $\mathcal{A}$. 
Since this is of the same order as the annulus one-point function,  
after dividing it by the disk one-point function and $g_s$,
we can interpret this as an additive contribution to $g$ proportional to $\mathcal{A}$, which we call $g_\text{ghost}$.

\begin{figure}[t!]
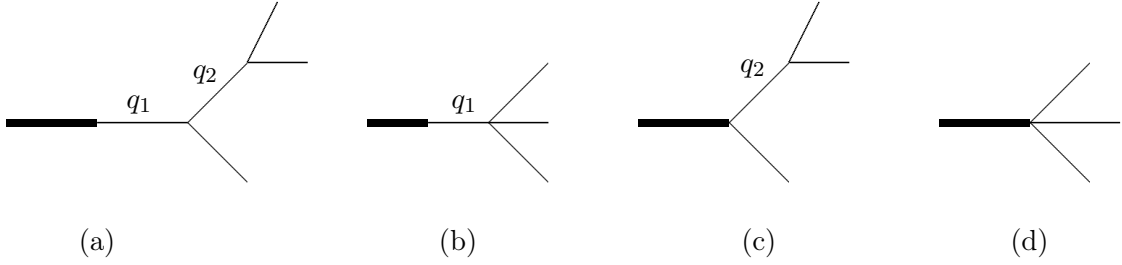

    \begin{center}
        \hbox{\figcooo}
    \end{center}
    \vspace{-1.2in}
    \caption{The four Feynman diagrams that contribute to the C-O-O-O amplitude.
    This is needed to evaluate the field-dependent relationship between the string field theory gauge parameter and the rigid gauge parameter that rotates the phase of an open string field with one end on the instanton.
    Figure adapted from \cite{Sen:2020eck}.}
    \label{fig:cooodiagrams}
\end{figure}

So, now we turn our attention to the computation of $g_\text{ghost}$.
The upper half plane C-O-O-O amplitude has two real moduli. 
We insert the cosmological constant operator at $z = \i$, and the three open string vertex operators $I$, $\frac{1}{2}c\partial c \partial^2 c$, $I$ are inserted at $z_1$, $z_2$ and $z_3$, respectively.
We only consider one cyclic ordering of $I$, $\frac{1}{2}c\partial c \partial^2 c$, $I$, which is the order in which we just wrote them.
The other cyclic ordering vanishes due to the trace over the Chan-Paton factors.
Up to permutations of the external legs,
there are four Feynman diagrams contributing to this amplitude; they are shown in figure \ref{fig:cooodiagrams}.

Let $\beta_1$ and $\beta_2$ be the coordinates on the moduli space, and denote $\vec{\beta} = (\beta_1, \beta_2)$.
This means that the locations $z_1$, $z_2$, $z_3$ of the open string punctures are functions of $\vec{\beta}$.
Let the local coordinates $w_1,w_2,w_3$ around the three punctures be related to the
UHP coordinate $z$ via
\begin{align}
    z = F_a(w_a,\vec{\beta}) = f_a(\vec{\beta}) + g_a(\vec{\beta})  w_a
    + \frac{1}{2} h_a(\vec{\beta}) w_a^2 + O(w_a^3)\, .
    \label{z_wa}
\end{align}
Then, by the general rules of \cite{Zwiebach:1992ie, Zwiebach:1997fe, deLacroix:2017lif}, we have
\begin{align}
    g_\text{ghost} = K_3 \int \d\beta_1 \hspace{-0.05in} \wedge  \hspace{-0.04in} \d\beta_2 
    \left\langle 
    \left\{ \sum_{a=1}^3 \oint \frac{\d z}{2\pi\i} \frac{\partial F_a}{\partial         \beta_1} b(z) 
    \right\}
    \left\{ \sum_{a=1}^3 \oint \frac{\d z}{2\pi\i} \frac{\partial F_a}{\partial         \beta_2} b(z) 
    \right\}
    \left\{ \frac{1}{2} c \partial c \partial^2 c (z_2)
    \right\} 
    \, c\cbar V(\i) 
    \right\rangle \, ,
    \label{gghost_def}
\end{align}
where we used the fact that all three open string insertions, being dimension zero primaries, are unaffected by the conformal transformations $F_a$.
The constant $K_3$ is a constant of proportionality that has been determined in appendix \ref{app:gaugeredefinition};
the result is $K_3 = - \frac{\i b}{\pi Q}$.\footnote{Integration over $\beta_1$ and $\beta_2$
includes integration over the Schwinger parameters of all the propagators appearing
in figure \ref{fig:cooodiagrams}, so unlike in the case of $K_1$ appearing in
\refb{gpsidef}, we do not need to
include any propagator factors in the definition of $K_3$.}
Using \refb{gghost_def} together with (\ref{z_wa}), the $bc$ OPE, and equations (\ref{vonepointi}), (\ref{cghostcorrelator}), one finds
\begin{align}
    g_\text{ghost} = - \frac{1}{4\pi} \int \left[ 
    (1+f_2^2)g_2^{-3} \d h_2 \wedge \d g_2 
    - 2 f_2 g_2^{-2} \d h_2 \wedge \d f_2 + 
    (2 f_2 h_2 g_2^{-3} + 2 g_2^{-1}) \d g_2 \wedge \d f_2\right]
    \, .
    \label{gghost2}
\end{align}
Note that only $(f_2, g_2, h_2)$ appear in the above expression since the insertions at $z_1$ and $z_3$ are the identity operator.

Next, we note that the two-form that appears in (\ref{gghost2}) is exact, and so $g_\text{ghost}$ can be written as
\begin{align}
    g_\text{ghost} &= -\frac{1}{2\pi} \int \d J \label{gghost_dj} \\
    J &:= - f_2 g_2^{-1} \d g_2 - \frac{1}{2} (1+f_2^2)h_2 g_2^{-3} \d g_2 + 
    \frac{1}{2}(1+f_2^2) g_2^{-2}\d h_2 + \d f_2 
    - \frac{2}{1+f_2^2} \, \d f_2\, .
    \label{jdef}
\end{align}
The final two terms vanish when we compute $\d J$, but they are necessary. 
The $\d f_2$ term has been chosen so that $J$ is invariant under the PSL$(2,\mathbb{R})$ transformation $z\to \frac{z-c}{1+c z}$ that keeps the closed string puncture at the point $z = \i$ fixed but moves the three open string punctures.
This ensures that there are no special contributions from the region of moduli where the open string punctures go to infinity along the real axis.
The final term in (\ref{jdef}) is PSL$(2,\mathbb{R})$ invariant by itself and has been chosen so that the contraction of $J$ with the vector $(\delta f_2, \delta g_2, \delta h_2) = (1+f_2^2, 2 f_2 g_2, 2 g_2^2 + 2f_2 h_2)$ that generates the above PSL$(2,\mathbb{R})$ transformation vanishes.
This final term was missing in the analysis of \cite{Sen:2020eck}. 
Without this term $J$ will
not be a well-defined one-form in the two-dimensional moduli space parametrized by
$\beta_1,\beta_2$, and we cannot apply Stokes's theorem to evaluate the right hand side
of \refb{gghost_dj} as a boundary term.

Let us elaborate a bit more about this point.
One could fix the PSL$(2,\mathbb{R})$ transformation above by working with local coordinates such that, say, $f_2 = 0$.
We can do this by making the transformation $\widetilde{z} = \frac{z - f_2}{1 + f_2 z}$.
Applying this transformation to (\ref{z_wa}), we find the new local coordinate around the second puncture to be
\begin{align}
    \widetilde{z} = \frac{g_2}{1+f_2^2}\, w_2 + 
    \frac{1}{2} \frac{h_2 + h_2 f_2^2 - 2 f_2 g_2^2}{(1+f_2^2)^2} \, w_2^2 + O(w_2^3) =: 
    \widetilde{g}_2 \, w_2 + \frac{1}{2} \widetilde{h}_2 \, w_2^2 + O(w_2^3)\, ,
    \label{g2tildeh2tildedef}
\end{align}
where the last equality serves to define the quantities $\widetilde{g}_2$ and $\widetilde{h}_2$.
In this `gauge', both the terms proportional to $\d \widetilde{f}_2$ in (\ref{jdef}) drop out and one gets $J = - \frac{1}{2} \, \widetilde{h}_2 \widetilde{g}_2^{-3} \d \widetilde{g}_2 + \frac{1}{2}\, \widetilde{g}_2^{-2} \d \widetilde{h}_2$.
Now, using the definitions of $\widetilde{g}_2$ and $\widetilde{h}_2$ from (\ref{g2tildeh2tildedef}), we can transform back to the $(f_2, g_2, h_2)$ variables, and one finds (\ref{jdef}).

The benefit of using (\ref{jdef}) is that since all the open string punctures are treated
in the same way, the three different orderings of the open string punctures on the real line (preserving the cyclic ordering) that need to be summed can be obtained by a simple permutation of the labels.
We have checked that using the gauge-fixed form of $J$ and summing over the three orderings gives the same final answer.

\begin{figure}[t!]
    \centering
    \includegraphics[width=0.5\textwidth]{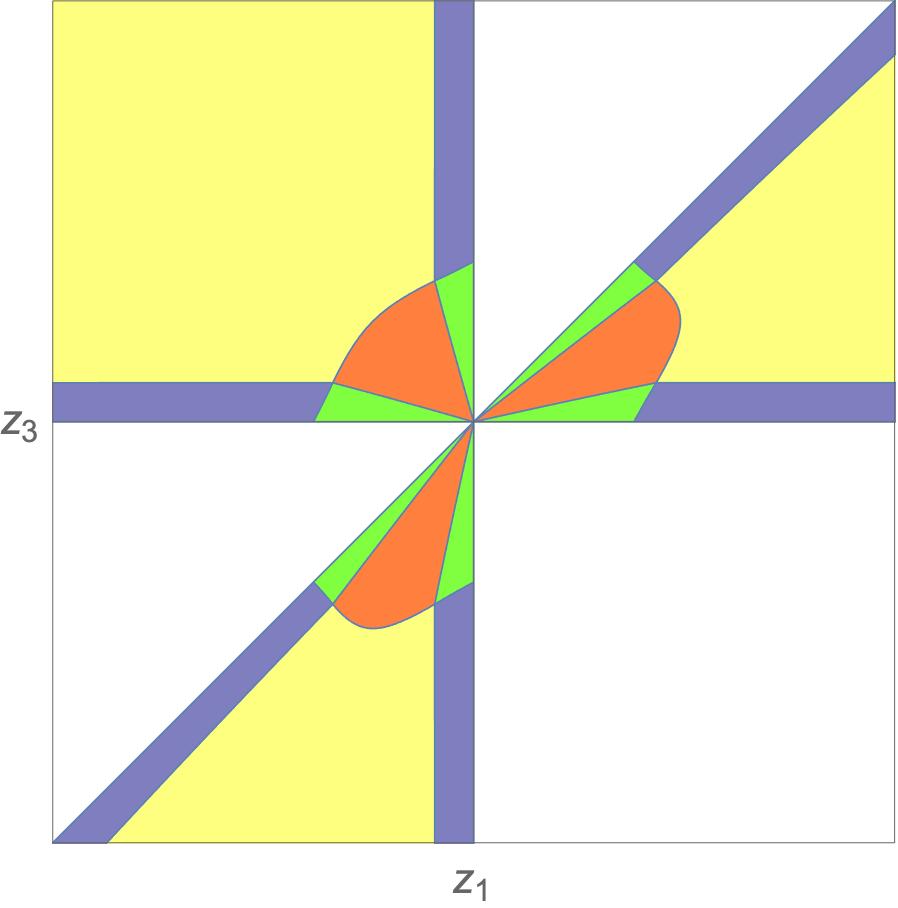}
    \caption{The regions of moduli space of the C-O-O-O amplitude (with a fixed cyclic ordering) corresponding to the four Feynman diagrams in figure \ref{fig:cooodiagrams}. 
    For this figure, we have set the location of the second puncture $z_2 = 0$.
    The green region corresponds to figure \ref{fig:cooodiagrams}(a), the red region to figure \ref{fig:cooodiagrams}(b), the blue region to figure \ref{fig:cooodiagrams}(c), and the bulk of the moduli space, shown in yellow, corresponds to figure \ref{fig:cooodiagrams}(d). The chosen cyclic ordering $(z_1, 0, z_3)$ gives rise to three linear orderings, namely, 
    $z_1 \leq 0 \leq z_3$, $0 \leq z_3 \leq z_1$ and $z_3 \leq z_1 \leq 0$, 
    which can be clearly identified. The same color is used to label different regions
    of the moduli space related by cyclic permutation of external open string states.}
    \label{fig:z1z3reions}
\end{figure}

We now turn to the computation of the four Feynman diagrams in figure \ref{fig:cooodiagrams}. 
The corresponding regions in moduli space are shown in figure \ref{fig:z1z3reions}.
Except for the overall normalization and the last term in \refb{jdef}, the analysis is identical
to that in appendix D of \cite{Sen:2020eck}. Therefore we shall be brief.

The Feynman diagrams in figure \ref{fig:cooodiagrams}(a) and \ref{fig:cooodiagrams}(c) vanish.
This is because, given the choice of the three open string vertex operators, neither the tachyon nor the out-of-Siegel gauge mode $\psi$ can propagate along the propagator labeled by $q_2$ in figure \ref{fig:cooodiagrams}.
The propagation of other $L_0 = 0$ Siegel gauge states are prevented by the replacement rule (\ref{replacementrule}), and contributions from the $L_0 > 0$ states are suppressed in the limit of large $\lambda$ and $\alpha$.

Since the integrand in (\ref{gghost_dj}) is a total derivative, we can evaluate the Feynman diagram corresponding to the bulk of moduli space, i.e. figure \ref{fig:cooodiagrams}(d), via Stokes's theorem.
This region shares one boundary with the region corresponding to figure \ref{fig:cooodiagrams}(b) (the yellow-red boundary in figure \ref{fig:z1z3reions}), 
and 
one boundary with the region corresponding to figure \ref{fig:cooodiagrams}(c) (the yellow-blue boundary in figure \ref{fig:z1z3reions}).
Let us denote these two contributions as $g_\text{ghost}^{(b)\mhyphen (d)}$ and $g_\text{ghost}^{(c)\mhyphen (d)}$, respectively.

Next, we observe that \cite{Sen:2020eck}
\begin{align}
    g_\text{ghost}^{(b)\mhyphen (d)} + g_\text{ghost}^{(b)} = 0 \, .
\end{align}
This is because if we denote the relevant part of the $\int\d J$ in region (b) as $\int q_1^{-2}\d q_1 \wedge f(\tau)\d \tau$, where $\tau$ is the modular parameter in the O-O-O-O vertex, then our choice of $J$ gives $\int J= \int  f(\tau)\d \tau$ from the (b)-(d) boundary.\footnote{This was verified in \cite{Sen:2020eck} for the choice of $J$ in which
the last term in \refb{jdef} was absent. We have checked that the extra term does not affect
the result. This is a consequence of the fact that,  for figure \ref{fig:cooodiagrams}(b),
all the open string vertex operators remain close to the origin of the UHP, and hence the
change in $f_2$ along the (b)-(d) boundary is small for large $\wt\lambda$ and $\alpha$.
}
Evaluating the contribution from figure \ref{fig:cooodiagrams}(b) using the replacement rule (\ref{replacementrule}), we get $-\int \d \tau f(\tau)$, which cancels the contribution 
of $\int J$ from the (b)-(d) boundary.
This is similar to the mechanism discussed in section \ref{sec:gws}.

Thus the only contribution to $g_\text{ghost}$ comes from $g_\text{ghost}^{(c)\mhyphen (d)}$.
This is the boundary between the yellow and blue regions of figure \ref{fig:z1z3reions}, which consists of six segments.
Each of these six segments lies at $q_2 =1$ and can be parametrized by the quantity $\beta$ that appears in the definition of the C-O-O vertex in section \ref{sec:vertices_review} in the range $\frac{1}{2\lt} \leq \beta \leq 1$.
Explicit expressions for $f_2$, $g_2$ and $h_2$ for each segment can be found in 
equations (D.24)-(D.26) of \cite{Sen:2020eck}.
Apart from the normalization constant discussed above, the only change compared to \cite{Sen:2020eck} is due to the final term $-\frac{2\d f_2}{1+f_2^2}$ in the one-form $J$ (\ref{jdef}).
This term precisely cancels the $O(1)$ contribution to $g_\text{ghost}$ in \cite{Sen:2020eck}. 
The conclusion is that
\begin{align}
    g_\text{ghost} &= g_\text{ghost}^{(c)\mhyphen (d)} = -\frac{1}{\pi} \cdot \lt  \int_{(2\lt)^{-1}}^1 \d\beta\, f'(\beta)
    = \frac{\lt}{2\pi}\, .
    \label{gghost}
\end{align}
We have added together the contribution from the six boundary segments and used (\ref{fvalues}) to get to the final expression.

For comparison, note that the result of \cite{Sen:2020eck} was $g_\text{ghost} = \frac{\lt}{2\pi} - \frac{1}{2}$.
With the modification to $J$ that we have discussed, the function $g(\omega)$ computed in \cite{Sen:2020eck} will have an extra contribution $+\frac{1}{2}$.
This precisely resolves the mismatch with the matrix model results of \cite{Balthazar:2019rnh, BRYunpublished}.

\subsection{Final result and concluding remarks}
Adding together the worldsheet contribution (\ref{gws}), the $\psi$-exchange contribution (\ref{gpsi}), and the contribution (\ref{gghost}), we find
\begin{align}
    g = g_\text{ws} + g_\psi + g_\text{ghost} = \frac{1}{2}\, ,
\end{align}
in perfect agreement with the general prediction (\ref{efgexp}).

As already mentioned at the end of section \ref{sec:gghost}, if we use the corrected form of $J$ in (\ref{jdef}) for the computation in \cite{Sen:2020eck} of the annulus one-point function in $c=1$ string theory, we also find a perfect match with the results of Balthazar, Rodriguez, and Yin \cite{Balthazar:2019rnh, BRYunpublished}.
In the notation of \cite{Sen:2020eck}, this would mean that $A_g = 0$.

We conclude this article with a remark about the quantity $C = \widetilde{C}$ in (\ref{eaafgexp}), which represents $O(g_s)$ corrections to the instanton action.
Corrections to this order have been computed in one-matrix integrals \cite{Marino:2007te, Marino:2008vx}.
It would be interesting to compute this quantity directly from the worldsheet of the $(2,p)$ minimal string and check that it agrees with the prediction from the dual one-matrix integral.
The main technical challenge seems to be that the moduli spaces of the relevant Riemann surfaces, the disk with a handle, and the three-holed sphere,  are three-dimensional,  which is one higher than the ones that we have analyzed.

\paragraph{Acknowledgments.} 
D.S.E. would like to acknowledge the Shoucheng Zhang Graduate Fellowship for support.
R.M.  is supported in part by AFOSR grant FA9550-16-0092,  Simons Investigator Award \# 508298,  and National Science Foundation grant PHY-2014215. 
P.M.  is supported by the Department of Atomic Energy, Government of India, under project no. RTI4001.
C.M. is supported in part by the U.S. Department of Energy, Office of Science, Office of High Energy Physics under QuantISED Award DE-SC0019380 and contract DE-AC02-05CH11231.
A.S. is supported by ICTS-Infosys Madhava Chair Professorship and the J. C. Bose fellowship of the Department of Science and Technology, India.

\appendix

\section{Fixing PSL$(2,\mathbb{R})$ with two closed string punctures}
\label{app:fixtwoclosedstrings}
We want to study the integration measure for the upper half plane amplitude $A(\psi_c^n\psi_o^m)$ when all the open string punctures are integrated, one closed string puncture is fixed at $\i$, and another closed string puncture is fixed at $\i y$.

The only non-trivial part is the overall normalization since the
$y$ dependence of the measure is captured by the correlation function where appropriate ghost factors are included in the definition of the unintegrated vertex operators.
We start with the original configuration with one open string vertex operator fixed at the origin and one closed string vertex operator fixed at $z = \i$. 
For convenience, let us consider the case where we have only one open string vertex operator and two closed string vertex operators.
We take the unintegrated form of the closed and open string vertex operators to be $\psi_c = c\cbar V_c$ and $\psi_o = c V_o$, respectively.
(Here $V_c$ is a dimension $(1,1)$ bulk operator, and $V_o$ is a dimension 1 boundary operator in the worldsheet theory.)
Let the integrated closed string vertex operator be at $z=x+\i y$. 
In this case the integrand is
\be \label{eap3}
{\d x \d y \over \pi} \, \frac{\i\pi}{g_s} \, \langle \psi_c(\i) V_c(z) \psi_o(0)\rangle \, .
\ee
Now we shall make a PSL$(2,\mathbb{R})$ transformation
\be
z' ={z - a\over 1+a z}\, .
\ee
Denoting $z' = x' + \i y'$, we adjust $a$ so that $x'$ vanishes.
This will move the open string at the origin to $x''=-a$. 
The new configuration is labelled by $x''$ and $y'$ and our goal will be to rewrite the original measure as
\be\label{eap4}
C \, g_s^{-1}\,  \d x'' \d y' \langle \psi_c(\i) (c + \cbar) V_c(\i y' ) V_o(x'')\rangle \, .
\ee

It is the constant $C$ that we want to determine. This can be
done by taking $(x,y)$ to be close to the origin so that $x'=0$ can be achieved
by an infinitesimal transformation with $a=x$. 
This gives $x''=-a=-x$ and $y'=y$. 
Therefore we have $\d x \d y = \d x'' \d y'$, and both $x'$ and $x''$ are integrated from $-\infty$ to $+\infty$.
Using (\ref{cghostcorrelator}), the ghost correlator in \refb{eap3} is
\be
\langle c\bar c(\i) c(0) \rangle = - 2\, \i\, ,
\ee
while that in \refb{eap4} is
\be
\langle c\bar c(\i) (c(\i y')+\bar c(\i y')) \rangle \approx - 4\, \i \quad\hbox{for $y'\approx 0$}\, .
\ee
Equality of \refb{eap3} and \refb{eap4} now gives
\be
C ={\i\over 2}\, . 
\ee
Even though we have derived this in the special case of only two closed strings and one open string, we can now generalize this to give the amplitude of $n$ closed strings and $m$ open strings, keeping fixed one closed string vertex operator at $z = \i$ and integrating another closed string vertex operator along the imaginary axis with measure $\d y$ from $y = 0$ to $y = 1$:
\be
A(\psi_c^n \psi_o^m) = \frac{\i}{2g_s} \int \langle \psi_c^n \psi_o^m \rangle_{\text{UHP}}\, .
\ee
This is the result quoted in (\ref{fixtwoclosed}).

\section{Tachyon exchange contribution to the disk two-point function}
\label{app:cancellation}

In this appendix we analyze the tachyon exchange contribution to the disk two-point function that was the subject of section \ref{sec:disk2pt}.
Our goal is to show that the tachyon exchange explicitly cancels the divergent term in (\ref{avvdiv}).

The open-closed interaction vertex is associated with an UHP two-point function, with one closed string at $\i$ and one open open string at $0$.
Let us denote by $z$ and $z'$ the coordinates on the two UHP coordinates representing the two vertices respectively. 
The closed strings are inserted at $z=\i$ and $z'=\i$, while the open strings are inserted at $z=0$ and $z'=0$.
Since the open string insertions are part of the internal open string propagator, they are in general off-shell and we need to specify the local coordinates $w$ and $w'$ around the insertion points $z=0$ and $z'=0$. 
We take \cite{Sen:2020eck}
\be\label{elocal}
w=\lambda z, \qquad w'=\lambda z'\, ,
\ee
for some large constant $\lambda$.
Then the string field theory Feynman diagram produces part of the world-sheet moduli space where we sew the two upper half planes by the relation
\be 
w w' = -q, \qquad 0\le q\le 1\, .
\ee
Using \refb{elocal} we get the relation
$z = -q/(\lambda^2 z')$,
so that the two closed-string vertex operators are inserted in the $z$-plane at $z = \i$ and $z = \i q / \lambda^2$.
Calling the second position $\i y$ we see that the range $0\le q\le 1$  translates to $0\le y\le \lambda^{-2}$. 
Since the desired range in section \ref{sec:disk2pt} is $0\le y\le \eps$, we see that we should choose
\be
\lambda^2 = 1/\eps\, .
\ee

Once we have determined the relation between the parameters $\eps$ and $\lambda$, we shall forget the Schwinger parameter representation of the propagator and directly calculate the contribution from various open string exchange diagrams. 
Due to the relation \refb{elocal} between the local and global coordinates, the coupling of an internal open string state of conformal weight $L_0 = h$ is scaled by a factor of $\lambda^{-h}$.
Taking into account that we have two upper half plane amplitudes connected by an open
string propagator, we get a net factor of $\lambda^{-2h}=\eps^h$. 

The leading contribution, associated with the exchange of the open string tachyon $c_1|0\rangle = c(0)|0\rangle$ with $L_ 0 = -1$, is given by,
\be\label{edivpart}
\eps^{-1}  \cdot 
    \frac{\i\pi}{g_s} \langle c\bar c V(\i) c(0)\rangle_{\text{UHP}} \, \cdot 
    {g_o^2 \over (-1)}\, \cdot 
    \frac{\i\pi}{g_s} \langle c\bar c V(\i) c(0)\rangle_{\text{UHP}}\, .
\ee
The two $\i\pi g_s^{-1} \langle c\bar c V(\i) c(0)\rangle_{\text{UHP}}$ factors 
are the disk correlators describing the open-closed interaction vertex, with conventions as in (\ref{fix1c1o}).
The factor of $g_o^2$ can be regarded either as part of the propagator in the convention in which $g_o^{-2}$ appears as an overall factor in the open string field theory action,
or from the two open-closed string vertices if we normalize the open string field so that  kinetic term has no $g_o$ dependence. 
The $-1$ in the denominator of the middle term is a reflection
of the open string tachyon having mass$^2=-1$. 
Using (\ref{vonepointi}) and (\ref{cghostcorrelator}) we have
\be
\langle c\bar c V(\i) c(0)\rangle_{\text{UHP}} = 
\frac{Q}{4b} \cdot (-2\i) =  - \i \, {Q\over 2b}\, .
\ee
Using the relationship $1/(2\pi^2 g_o^2) = 1/g_s$ \cite{Sen:1999xm, Sen:1999nx, Schnabl:2005gv, Erler:2019fye}, we can simplify \refb{edivpart} to
\be\label{esftcont}
- \eps^{-1} \, g_s^{-1} \, \frac{Q^2}{8b^2} \, .
\ee
So we conclude that the exchange of the open string tachyon precisely cancels the divergent contribution in (\ref{avvdiv}).

Finally, we note that we could work with a more general choice of local coordinates $w,w'$ instead of (\ref{elocal}), for instance $w = \frac{\lambda z}{1-\gamma z}, w' = \frac{\lambda z'}{1-\gamma z'}$, for some constant $\gamma$.
In this case,  the tachyon exchange contribution will be modified, but there will also be an extra contribution from the exchange of the out-of-Siegel-gauge mode $\psi$ multiplying the state $\i c_0\ket{0}$. 
As in \cite{Sen:2020eck}, the sum of the two contributions, expressed as a function of $\epsilon$, is independent of the choice of local coordinates $w,w'$.

\section{Normalization of the worldsheet contribution to the annulus one-point function}
\label{sec:annulus_overall_constant}

The goal of this appendix is to derive the proportionality constant in (\ref{g_formula_1}), and also to write precise expressions for the small-$v$ and small-$x$ behavior of the integrand.

Recall that we can write $g_s g$ as 
\begin{align}
    g_s\, g &= \int_0^1 \d v \int_{0}^\frac{1}{4} \d x \, F(v,x)\, ,
\quad \text{with}  \label{c1}\\
    F(v,x) &= C\,  \Tr\left[V(w,\bar w) \, b_0 \, c_0
\, v^{L_0-1}\right]\, ,  \quad w := 2\pi(x+\i y)\, , \label{c2}
\end{align}
where $C$ is the normalization constant that we want to determine. 
Recalling the argument near (\ref{gd_def_dx2phi}), we can also write $F(v,x)$ as
\begin{align}
	F(v,x) &= \partial_x G(v,x) \, , \quad 
     G(v,x) := \frac{C}{16\pi^2 b} 
    \Tr\left[ \partial_x \phi(w, \wbar) \, b_0 \, c_0\, v^{L_0-1}\right]\, .
    \label{c3}
\end{align}
Our strategy for determining $C$ will be to compare the integrand to the contribution from the Feynman diagram in figure \ref{fig:worldsheetdiagrams} (a) for small $v$ and small $x$.
We would also like to compute $G(v,x)$ and $F(v,x)$ in the regions of small $v$ or small $x$ (with the second variable being not necessarily small).

Recall that $w$ is the coordinate on the strip $0 \leq \Re{w} \leq \pi$, with $\Re{w} = 2\pi x$. 
We can map this to the upper half plane by the map $z=e^{\i w}$.
The annulus is obtained from this via the identification $z \equiv v z$.
The transformation of $\partial \phi$ is given in (\ref{eftrs1}), with a similar transformation for $\overline{\partial}\phi$.
We get
\begin{align}
    \partial_w \phi(w,\wbar) &= \i \frac{Q}{2} + \i z\, \partial_z\phi(z,\zbar) 
    \label{dwdz1} \, , \\ 
    \partial_{\wbar} \phi(w,\wbar) &= -\i \frac{Q}{2} - \i \zbar\, \partial_{\zbar}\phi(z,\zbar) 
    \, . \label{dwdz2}
\end{align}
Thus, for small $x$, we can use the bulk-boundary OPEs (\ref{bulkbdryope1}), (\ref{bulkbdryope2}) to get
\begin{align}
    \partial_x \phi(w,\wbar) = 2\pi (\partial_w \phi(w,\wbar) + \partial_{\wbar} \phi(w,\wbar))
    =  2\pi \i \left(z \partial_z \phi(z,\zbar) - \zbar \partial_{\zbar} \phi(z,\zbar) \right)
    \approx - \frac{Q}{x}\, .
\end{align}
The definition of $G(v,x)$ in (\ref{c3}) now gives
\begin{align}
G(v,x) &\approx - C\, \frac{Q}{b} \, {1\over 16 \pi^2 x}\,
\Tr\left[b_0 \, c_0\, v^{L_0-1}\right] = -
C\, \frac{Q}{b} \, {1\over 16 \pi^2 x}\, 
\frac{Z(v)}{v} \, , \label{egvxone}\\
F(v,x) &\approx C \, \frac{Q}{b} \, \frac{1}{16\pi^2 x^2} \, \frac{Z(v)}{v} \, . \label{efvxone}
\end{align}
Here $Z(v)$ is the annulus partition function, given in (\ref{zofv}).

Next, we consider the small-$v$, finite-$x$ region. 
We can evaluate this correlation function by separating out the matter and ghost contributions. 
In the ghost sector, the leading and subleading contributions to the trace are as in (\ref{annulus-small-v}).
In  the matter sector the leading and subleading contributions for small $v$ come from just the vacuum state $|0\rangle$. 
Therefore
\begin{align}
    \Tr\left[ \partial_x \phi(w, \wbar) \, b_0 \, c_0\, v^{L_0-1}\right]
    \approx (v^{-2}-2v^{-1}+\OO(1)) \, 
    \langle 0| \partial_x \phi (w,\bar w) | 0\rangle\, \qquad \hbox{for small $v$}\, .
    \label{dxphi_annulus_smallv}
\end{align}
Since we do not have a trace in the Liouville sector, we no longer need the identification $z \equiv vz$.
Using (\ref{dwdz1}) and (\ref{dwdz2}), and $\langle \partial_z \phi(z,\zbar) \rangle_\text{UHP} = -\frac{Q}{z-\zbar}$ as given in (\ref{ephiuhp}),  we get
\begin{align}
    \langle 0| \partial_x \phi (w,\bar w) | 0\rangle = 
    -\i \, 2\pi Q \, \frac{z+\zbar}{z - \zbar} 
    = - 2\pi Q \, \cot (2\pi x)\, .
    \label{dxphi_smallv}
\end{align}
Thus, from (\ref{c3}), we get
\begin{align}
G(v,x) &\approx - C \, \frac{Q}{8\pi b} (v^{-2} - 2 v^{-1} + \OO(1)) \cot(2\pi x) \, , \label{ecco4}\\
F(v,x) &\approx C \, \frac{Q}{b} \, (v^{-2}-2v^{-1}
+\OO(1)) \, 
{1\over 4\sin^2(2\pi x)} \quad \hbox{for small $v$}\, . \label{ecco3}
\end{align}

We shall now fix the constant $C$  by comparing (\ref{efvxone}) or (\ref{ecco3}) with the result $g^{(a)}$ of the Feynman diagram in figure \ref{fig:worldsheetdiagrams}(a), which corresponds to small $v$ and small $x$.
For this, we first rewrite $F(v,x)$ in the $(q_1,q_2)$ coordinates using eqs (4.73) and (4.81) of \cite{Sen:2020eck} for small $v$ and small $x$:
\be
v \approx {q_2 \over \alpha^2} , \qquad x = {q_1\over 2\pi \wt\lambda}\, .
\ee
This yields
\be\label{eccomp1}
F(v,x)\, \d v \, \d x \approx 
C \, \frac{Q}{b} \, \frac{1}{16 \pi^2} \, 
\frac{\d v \, \d x}{v^2\, x^2}
\approx
C \, \frac{Q}{b} \, \frac{\widetilde{\lambda}\alpha^2}{8 \pi} \,
 {\d q_2 \over q_2^2} \,  {\d q_1\over q_1^2}\, .
\ee

The Feynman diagram of figure \ref{fig:worldsheetdiagrams}(a) can be analyzed as follows. 
The leading contribution for small $v$ and small $x$ comes from the propagation of the open string tachyon along both propagators, producing the factor $\int \d q_2\, q_2^{-2} \, \d q_1 \, q_1^{-2}$. 
Since the tachyon vertex operator $c$ is a primary of dimension $-1$, the open-closed and open-open-open interaction vertices produce factors of $\lambda$ and $\alpha^3$, respectively. 
The three open string vertex produces a factor of the open string coupling constant $g_o$. 
The ratio of the open-closed vertex (\ref{fix1c1o}) to the disk one-point function (\ref{adiskv}) is given by
\be\label{ebb13}
 g_o \, { \i \pi \, \langle c(0) c\bar c(\i)\rangle_{\text{UHP}}\over {1\over 4} 
\langle (\p c - \bar\p\bar c) c\bar c(\i)\rangle_{\text{UHP}}} = \pi g_o\, .
\ee
The factor of $g_o$ arises due to the presence of the extra open string.
We have evaluated the ghost correlators using (\ref{cghostcorrelator}).
Putting all the factors together we get the following result for $g_s g^{(a)}$:
\be\label{eccomp2}
g_s g^{(a)} = \int\int {\d q_2\over q_2^2}\,  {\d q_1 \over q_1^2} \, 
\lambda \, \alpha^3\, \pi \, g_o^2 \, 
= \int\int{\d q_2\over q_2^2}\,  {\d q_1 \over q_1^2}
\frac{\lt \alpha^2}{2\pi} \, g_s \, .
\ee
In going to the second expression, we have used $\lt = \lambda \alpha$, and also the fact that $1/(2\pi^2 g_o^2) = 1/g_s$ \cite{Sen:1999xm, Sen:1999nx, Schnabl:2005gv, Erler:2019fye}.
Comparing \refb{eccomp1} and \refb{eccomp2}, we get,
\be
C = \frac{4 b}{Q} \, g_s\, .
\label{cresultannulus}
\ee
Plugging this value of $C$ into (\ref{c2}) and (\ref{c1}), we get the desired result (\ref{g_formula_1}).

For later use, we also rewrite (\ref{egvxone}), (\ref{efvxone}), (\ref{ecco4}) and (\ref{ecco3}) using (\ref{cresultannulus}):
\begin{align}
G(v,x) &\approx \begin{cases}
{-g_s\over 2\pi} \,  \cot(2\pi x) \, \{v^{-2}-2v^{-1}+\OO(1)\} \quad 
\hbox{for small $v$} \\
{-g_s\over 4 \pi^2 x}  \frac{Z(v)}{v}\quad\quad \hspace{1.52in}
\hbox{for small $x$}\, .
\end{cases}
\label{eGexp} \\
F(v,x) &\approx \begin{cases}
{g_s\over \sin^2(2\pi x)} \, \{v^{-2}-2v^{-1}+\OO(1)\} \quad 
\hbox{for small $v$} \\
{g_s\over 4 \pi^2 x^2}  \frac{Z(v)}{v}\quad\quad \hspace{1.12in}
\hbox{for small $x$}\, .
\end{cases}
\label{efvxtwonew}
\end{align}

\section{Normalization of the $\psi$ exchange contribution}
\label{app:psinormalization}
In this appendix we want to compute the normalization constant appearing in (\ref{gpsidef}).

The strategy to determine $K_1$ is to compute the Feynman diagram in figure \ref{fig:worldsheetdiagrampsi}(b), but with the tachyon running in the loop instead of the $\psi$ field, and compare the result with (\ref{gc}). 
Recall that figure \ref{fig:worldsheetdiagrampsi}(b) corresponds to $\frac{1}{2\lt} \leq \beta \leq 1$, in which the two open string punctures are not close to each other.
Evaluating the Feynman diagram in figure \ref{fig:worldsheetdiagrampsi}(b) with the tachyon in the open string loop requires evaluating the right hand side of (\ref{gpsidef}) with two $c$ operators instead of the two $\i \partial c$ operators.
The propagator of the tachyon field equals $-1$ in our normalization as opposed
to $1/2$ for the $\psi$ propagator. This yields
\begin{align}
    g^{(c)} 
    &=  2\, K_1 \int_{(2\lt)^{-1}}^1 \d\beta \sum_{a=1}^2 \oint_a \frac{\d z}{2\pi \i} 
    \frac{\partial F_a}{\partial \beta} \, \langle 
    b(z) F_1 \circ c(0) \, F_2 \circ c(0) \, c\cbar V(\i)
    \rangle_\text{UHP} \\
    &=  2\, K_1 \, \frac{\i Q}{b} \, \alpha^2 \lt^2 \left(1 + \frac{1}{4\lt^2} \right)^2
    \int_{(2\widetilde{\lambda})^{-1}}^1 \frac{\d\beta}{1+\beta^2}\, . \label{d3}
\end{align}
To get to the second line, we used the $bc$ OPE, the explicit form of $F_a$ given in (\ref{f1wbeta}), (\ref{f2wbeta}), and the correlators (\ref{cghostcorrelator}), (\ref{vonepointi}).
Comparing (\ref{d3}) to the first expression in the second line of (\ref{gc}), we find
\begin{align}
    K_1 = \frac{\i b}{2\pi Q} \, .
\end{align}
We can find the same result if we study the figure \ref{fig:worldsheetdiagrampsi}(a) with the tachyon running in the loop and comparing the result to (\ref{ga}).

\section{Normalization of the C-O-O-O amplitude}
\label{app:gaugeredefinition}

In this appendix, we determine the normalization constant $K_3$ appearing in (\ref{gghost_def}).
Our strategy to determine $K_3$ is to study the Feynman diagram in figure \ref{fig:cooodiagrams}(c), but with all three open string insertions being the tachyon field, with Chan-Paton factors kept the same as in section \ref{sec:gghost}.
The consistency condition between two different ways of computing this diagram will yield the value of $K_3$.
Let us denote the amplitude computed using \refb{gghost_def} with all the external
states taken
as $c$ by $\widetilde{A}$, and let $\widetilde{A}_c$ denote the contribution to this amplitude from figure \ref{fig:cooodiagrams}(c) (with just one of the three possible assignments of the external operators to the external legs).

First, note that figure \ref{fig:cooodiagrams}(c) with the internal propagator also being the tachyon, is the product of three factors: the C-O-O amplitude (with the closed string puncture being $V$ and the two open string punctures being tachyons), the tachyon propagator, and the O-O-O three tachyon amplitude.
The product of the C-O-O amplitude and the tachyon propagator appears also in figure \ref{fig:worldsheetdiagrams}(c) and is given by (\ref{gc}).\footnote{Note that \refb{gc}
includes division by $g_s$ and the disk one point function; so the normalization of the
amplitude determined this way directly yields the contribution to $g$.}
The O-O-O tachyon vertex can be computed using the local coordinates in (\ref{ooo-w}) and the ghost correlator in (\ref{cghostcorrelator}) to be $\alpha^3$.
So we find
\begin{align}
    \widetilde{A}_c = 
    - \frac{1}{\pi} \, \lt^2 \alpha^5 \left(1 + \frac{1}{4\lt^2} \right)^2 
    \int_{(2\lt)^{-1}}^{1} \frac{\d\beta}{1+\beta^2} \, .
    \label{atildec-1}
\end{align}

The second way of computing this amplitude is to start with the expression analogous to (\ref{gghost_def}).
Using $F_a \circ c(0) = g_a^{-1} c(z_a)$, we get the following expression:
\begin{align}
    \widetilde{A} &= 
    K_3 \int \d\beta_1 \hspace{-0.05in} \wedge  \hspace{-0.04in} \d\beta_2 
    \left\langle 
    \left\{ \sum_{a=1}^3 \oint \frac{\d z}{2\pi\i} \frac{\partial F_a}{\partial         \beta_1} b(z) 
    \right\}
    \left\{ \sum_{a=1}^3 \oint \frac{\d z}{2\pi\i} \frac{\partial F_a}{\partial         \beta_2} b(z) 
    \right\}
    (g_1 g_2 g_3)^{-1}
    c(z_1) c(z_2) c(z_3)
    \, c\cbar V(\i) 
    \right\rangle \nonumber \\
    &= - \frac{\i Q}{2b} K_3 \int 
    (g_1 g_2 g_3)^{-1} [ 
    (1+f_3^2)\, \d f_2 \wedge \d f_1  
    + (1+f_1^2)\, \d f_3 \wedge \d f_2
    + (1+f_2^2)\, \d f_1 \wedge \d f_3  
    ]\, .
\end{align}
We used the $bc$ OPE and equations (\ref{cghostcorrelator}), (\ref{vonepointi}),
\refb{z_wa} to get to the second line.
Let us use this expression to evaluate $\widetilde{A}_c$.
Using the explicit form of $f_a$, $g_a$ and $h_a$ suitable for the region corresponding to figure \ref{fig:cooodiagrams}(c) given in equation (D.24) of \cite{Sen:2020eck}, we find
\begin{align}
    \widetilde{A}_c &= \frac{\i Q}{2b} K_3\cdot   
    2\lt^2 \alpha^5 \left(1 + \frac{1}{4\lt^2} \right)^2
    \int_{(2\lt)^-1}^1 \frac{\d \beta}{1+\beta^2}
    \int_0^1 \frac{\d q_2}{q_2^2}
    \\
    & \to - \frac{\i Q}{b} K_3\,  
    \lt^2 \alpha^5 \left(1 + \frac{1}{4\lt^2} \right)^2
    \int_{(2\lt)^-1}^1 \frac{\d \beta}{1+\beta^2} \, ,
    \label{atildec-2}
\end{align}
where we used the replacement rule $\int_0^1 \frac{\d q_2}{q_2^2} \to -1$ for the plumbing fixture variable $q_2$ appearing in the internal propagator in figure \ref{fig:cooodiagrams}(c).

Comparing (\ref{atildec-1}) and (\ref{atildec-2}) we get
\begin{align}
    K_3 = - \frac{\i b}{\pi Q}\, .
\end{align}

\bibliographystyle{apsrev4-1long}
\bibliography{draft_mismatch}
\end{document}